\documentclass[12pt]{article}
\usepackage{graphics,color} %
\def\eq#1\en{\begin{equation} #1\end{equation}}
\def\pp#1{\partial_{#1}}
\begin{document}


\begin{titlepage}
\begin{flushright}
\begin{tabular}{l}
CERN-TH/2002-308\\
hep-ph/0212309\\
November 2002
\end{tabular}
\end{flushright}


\begin{center}
\boldmath
{\Large \bf {Rare and forbidden decays}$^*$}
\unboldmath


\smallskip
\begin{center}
{\sc {\large Josip Trampeti\'{c}}}\\
\vspace*{1mm}
{\sl Theory Division, CERN, CH-1211 Geneva 23, Switzerland\\
Theoretical Physics Division, Rudjer Bo\v skovi\' c Institute, Zagreb, Croatia}
\end{center}

{\large\bf Abstract\\[10pt]} \parbox[t]{\textwidth}{
In these lectures I first cover radiative and semileptonic B decays, including the 
QCD corrections for the quark subprocesses. The exclusive modes and the evaluation of the hadronic
matrix elements, i.e. the relevant hadronic form factors, are the second step. 
Small effects due to the long-distance, spectator contributions, etc.
are discussed next. The second section we started with non-leptonic decays, typically 
$B \to \pi\pi,\;K\pi,\; \rho\pi,...$ 
We describe in more detail our prediction for decays dominated by the $b\to s \eta_c$ transition.
Reports on the most recent experimental results are given at the end of each subsection.

In the second part of the lectures I discuss decays forbidden by the Lorentz and gauge invariance, and 
due to the violation of the angular moment conservation, generally called 
the Standard Model-forbiden decays.
However, the non-commutative QED and/or non-commutative Standard Model (NCSM),
developed in a series of works in the last few years allow some of those decay modes. 
These are, in the gauge sector, $Z\to \gamma\gamma,\; gg$, and 
in the hadronic sector, flavour changing decays of the type $K\to \pi \gamma$, $B\to K\gamma$, etc.
We shall see, for example, that the flavour changing decay $D^+_s\to \pi^+ \gamma$ dominates over 
other modes, because the processes occur via charged currents, 
i.e. on the quark level it arises from the point-like 
photon $\times$ current $\times$ current interactions. 
In the last section we present the transition rate of ``transverse plasmon'' 
decay into a neutrino--antineutrino pair via noncommutative QED, i.e. $\gamma_{\rm pl}\to \nu\bar\nu$.
Such decays gives extra contribution to the mechanism for the energy loss in stars.
}

\vspace{0.1cm}

{\sl $^*$Based on presentations given at the XLII Cracow School of \\Theoretical Physics,
     Zakopane, Poland, 31 May -- 9 June 2002;\\  
     and LHC Days in Split, Croatia,  8 - 12 October 2002.
     \\ Acta Physica Polonica {\bf B33}, 4317 (2002).
     }
\end{center}

\end{titlepage}

\thispagestyle{empty}
\vbox{}
\newpage

\setcounter{page}{1}


\newpage
\section{\it Rare B meson decays: theory and experiments}

\subsection{\it Introduction to the rare B meson decays}

The experimental challenge of finding new physics in direct
searches may still take some time if 
new particles or their effects set in only at several hundred GeV. Complementary to 
these direct signals at highest available energies are the measurements of the effects of 
new ``heavy'' particles in loops, through either precision measurements or detection
of processes occurring only at one loop in the Standard Model (SM).

In the light quark system, however, the presence of quite large long-distance (LD) effects that cannot be 
calculated reliably makes this study difficult, except in the extremely rare process $K \to \pi {\bar\nu}{\nu}$.
The situation is much better in the $b$ quark system. Among these are the transitions induced by 
flavour-changing neutral currents (FCNC). 
Rare decays of the B meson offer a unique opportunity to study  electroweak theory in higher orders.
Processes such as $b \to s \gamma$, $b \to s {\ell}^+{\ell}^-$ and $b \to s g$ do not occur at tree level, 
and at one loop they occur at a rate small enough to be sensitive to physics beyond the SM. 

Studying B meson radiative decays $B\to K^* \gamma$ based on the $b \to s \gamma$ quark transition, 
described by a magnetic dipole operator, 
we have found two major effects \cite{des}:\\
(1) Large QCD correction due to the introduction of 1-gluon exchange. 
One might say that 1-gluon exchange 
changes the nature, i.e. the functional structure of the GIM cancellation \cite{des,bbm}:
$(m_t^2-m_c^2)/m_w^2 \rightarrow \ln(m_{t}^2/m_{c}^2)$. 
Note, however, that since $m_{\rm top}\simeq 2m_W$, the GIM mechanism is no longer crucial 
and QCD corrections become modest.\\
(2) Huge recoil effect caused by the motion of the hadron as a whole producing a large suppression of 
the hadronic form factor \cite{des}.

To simplify the very first attempt of calculating the $b\to s \gamma$ and $B\to K^* \gamma$,
we have made few very important assumptions, which all turn 
to be right and proved within the past decade by a
number of authors. They become major advantages for studies of rare B meson decays:\\
(i) the B meson is made of a sufficiently heavy b quark, thus permitting the use of the spectator approximation in the 
calculation;\\
(ii) absence of large long-distance effects;\\
(iii) the $b \to s \gamma$ transition is the only contribution to $B^0$ decay;\\
(iv) the B meson lifetime is, relatively speaking, prolonged more than that of kaons
because of the smallness of $V_{cb}$ and $V_{cu}$,
allowing ${\bar B}B$ mixing to be studied.

Other important impacts of studies of B mesons are:\\
(I) tests of electroweak theory (SM) in one loop are of interest in their own right, because they verify
the gauge structure of the theory;\\
(II) the realization of a heavy quark symmetry, i.e. the structure of hadrons, becomes independent of flavour and spin 
(spin symmetry) for $1/m_b \to \infty$;\\
(III) the emergence of the Heavy Quark Effective Field Theory (HQET);\\
(IV) if there exists an enhancement of the SUSY over SM contribution, it is clear that the B meson radiative processes, 
dominated by the $b \to s \gamma$ quark 1-loop transition, can be an interesting candidate 
directly affected by the SUSY contribution \cite{bbm1};\\
(V) the decay $b \to s \gamma$ is by far the most restrictive process 
in constraining the parameters of the charged Higgs boson
sector in 2 Higgs doublets model, yielding bounds that are stronger than those from other low-energy processses and from 
direct collider searches \cite{joan}.

Today allmost everybody in the particle physics community agrees that B decays in general do become one of the
most important classes of tests of the SM and physics beyond the SM.

Although the quark level calculations are fairly precise in the $b$ quark system, one is still hampered
by the lack of knowledge of the hadronic form factors. However, in the past decade there has been extensive activity
in the form factor evaluation using the perturbative QCD technique with the help of the HQET and from 
improving lattice model calculations.

The first observations of the exclusive $B\to K^* \gamma$ decays were reported in 1993/94 by the 
CLEO Collaboration \cite{cl}.

On the experimental side the last two years were especially exciting since BaBar and Belle Collaborations joined
CLEO Collaboration in producing and publishing a large number of data concerning the B meson decays.

\subsection{\it Radiative and semileptonic B decays}

\begin{figure}
 \resizebox{0.9\textwidth}{!}{%
  \includegraphics{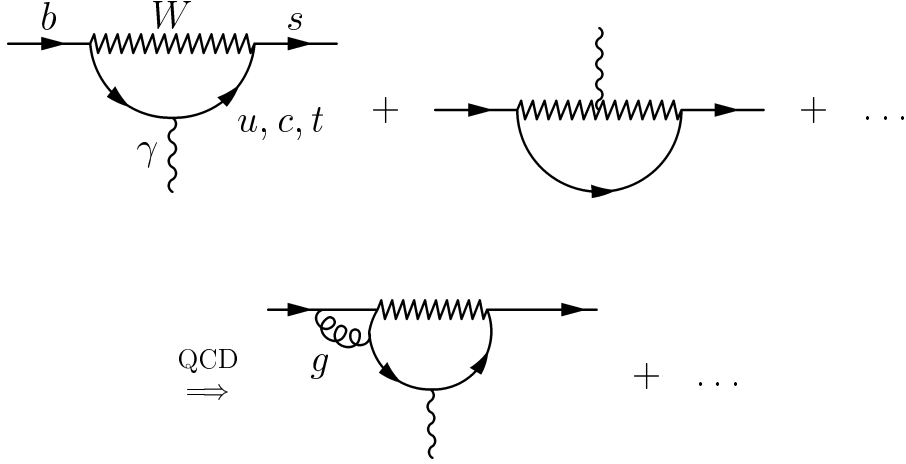}}
 \caption{The penguin diagrams, including the QCD short-distance corrections, 
 contributing to the $b\to s \gamma$ transition.}
 \label{Zfig1}
\end{figure}

The $b \to s \gamma$ decay is a one-loop electroweak process that arises from the so-called penguin diagrams
through the exchange of $u$,$c$,$t$ quarks and weak bosons, see Fig. \ref{Zfig1},
 and is given by
\begin{eqnarray}
J_{\mu}= \left\{G_1 {\bar s}(\gamma_{\mu}q^2-q_{\mu}{\not \!q})b_L +
iG_2\left[ m_s{\bar s}\sigma_{\mu\nu}q^{\nu}b_L + m_b{\bar s}\sigma_{\mu\nu}q^{\nu}b_R\right]\right\},
\end{eqnarray}
where the first term for real photon vanishes identically, owing to the electromagnetic gauge condition. 
Using the standard parametrization of the CKM matrix in the case of three doublets, $G_2$ is given by
\begin{eqnarray}
G_2= \frac{G_F}{\sqrt 2}\frac{e}{4\pi^2}{\sum_{i=u,c,t}}A_i F_2(x_i);\;\; A_i=V^*_{is}V^{}_{ib},\;\; x_i=m^2_i/m^2_W,
\end{eqnarray}
were the modified Inami--Lim function $F_2(x_i)$ derived from the penguin (1-loop) diagrams is \cite{lim}
\begin{eqnarray}
F_2(x_i)= \frac{x_i}{12(1-x_i)^4}\left[ (1-x_i)(8x^2_i + 5x_i - 7)- 6x_i(2-3x_i)\ln x_i\right].
\end{eqnarray}
Introduction of 1-gluon exchange (QCD corrections) in penguin diagrams removes the power suppression,
i.e. $(m_t^2-m_c^2)/m_W^2 \rightarrow \ln(m_{t}^2/m_{c}^2)$; or one can say that QCD corrections
change the nature of the GIM cancellation from quadratic to logarithmic \cite{des,bbm}.
These QCD corrections also strongly affect the semileptonic transitions \cite{dt1}.
The following properties are important to note:\\
(a) the dominant contribution to the perturbative $b \to s \gamma$ amplitude originates from charm-quark loops;\\
(b) after inclusion of the QCD corrections, the top-quark contribution is less than 50 \% of charm and it comes 
with opposite sign;\\
(c) the up-quark contribution is suppressed with respect to charm by \\
$|V^*_{us}V^{}_{ub}/V^*_{cs}V^{}_{cb}| \simeq 2$\%.\\
It is necessary to consider the above facts when one attempts to extract 
the CKM matrix element $|V_{ts}|$ from $b \to s \gamma$.

Before proceeding, it is important to note that, 
{\it in the limit $m_q\to \infty$ the inclusive meson decay partial width is equal
to the free quark decay partial width}. In the case of B mesons the $b$ quark is 
sufficiently heavy to satisfy the above statement:
\begin{eqnarray}
\Gamma (B \to X_q \gamma,\; \ell^+\ell^-)|_{\rm inclusive \atop decay}=
\Gamma (B \to X_q \gamma,\; \ell^+\ell^-)|_{\rm free \;quark \atop decay}
\end{eqnarray}
where $q$ represents the light quarks.

\subsubsection{{\it Complete QCD corrected weak Hamiltonian density}}

In the SM, B decays are described by the effective weak Hamiltonian obtained
by integrating out heavy, i.e. the top-quark, W-boson and Higgs fields:
\begin{equation}
H_{\Delta B=1}^{\Delta S=-1}= 2\sqrt{2}G_F 
\left[\sum_{q=u,c} V_{qs}^*V_{qb}^{}(c_1O^q_1+c_2O^q_2)-V_{ts}^*V_{tb}^{}\sum_{i=3}^{10}c_iO_i\right]\;+\; {\rm h.c.} 
\end{equation}
The $O_i$'s are operators
\begin{eqnarray}
O^q_1&=&\left({\bar s}^i_L \gamma_{\mu} q^j_L\right)\left({\bar q}^j_L \gamma^{\mu}b^i_L\right),\;\;
O^q_2=\left({\bar s}^i_L \gamma_{\mu} q^i_L\right)\left( {\bar q}^j_L \gamma^{\mu}b^j_L\right),
\nonumber \\
O_{3 \choose 5}&=&\left({\bar s}^i_L \gamma_{\mu} b^i_L\right) 
\sum_{q'}{\bar q}^{\prime j}_{L \choose R} \gamma^{\mu}q^{\prime j}_{L \choose R},\;\;
O_{4 \choose 6}=\left({\bar s}^i_L \gamma_{\mu} b^j_L\right) 
\sum_{q'}{\bar q}^{\prime j}_{L \choose R} \gamma^{\mu}q^{\prime i}_{L \choose R},
\nonumber \\
O_7&=&\frac{e}{16\pi^2}\,m_b \,\left({\bar s}^i_L\sigma_{\mu\nu}b^i_R\right)\, F^{\mu\nu},\;\;
O_8=\frac{g}{16\pi^2}\,m_b \,\left({\bar s}^i_L\sigma_{\mu\nu}T^{ij}_ab^i_R\right)\, G_a^{\mu\nu},
\nonumber \\
O_9&=&\frac{e^2}{16\pi^2}\,\left({\bar s}^i_L\sigma_{\mu\nu}b^i_R\right){\bar\ell}\gamma^{\mu}\ell,\;\;
O_{10}=\frac{e^2}{16\pi^2}\,\left({\bar s}^i_L\sigma_{\mu\nu}b^i_R\right){\bar\ell}\gamma^{\mu}\gamma_5\ell,
\end{eqnarray}
where $F^{\mu\nu}$ and $G_a^{\mu\nu}$ are the electromagnetic and gluon interaction field strength tensors, respectively,
and $e$ and $g$ are the corresponding coupling constants.
The $c_i$'s are the well-known Wilson coefficients first calculated up to the next-to-leading order (NLO)  
in Ref. \cite{ciuc}. The calculation was performed with the help of the renormalization group 
equation whose solution requires the knowledge of the anomalous dimension matrix to a given order in 
$\alpha_s$ and the matching conditions: 
\begin{eqnarray}
c_1 (\mu)&=&\frac{1}{2}\left(\xi^{6/23} - \xi^{-12/23}\right),\;\; c_2 (\mu)=\frac{1}{2}\left(\xi^{6/23} + \xi^{-12/23}\right),
\nonumber \\
c_7(\mu)&=& \xi^{-16/23}\left[c_7(m_W)-\frac{8}{3}c_8(m_W)\left(1-\xi^{2/23}\right)+\frac{232}{513}\left(1-\xi^{19/23}\right)\right],
\nonumber \\
c_9(\mu)&=&c_9(m_W)-[c_1(\mu)+3c_2(\mu)]\ln\left[(m_c/\mu)^{-8/9}\right]
\nonumber \\
&-&\frac{4\pi}{\alpha_s (m_W)}\left[\frac{4}{33}\left(1-\xi^{-11/23}\right)-\frac{8}{87}\left(1-\xi^{-29/23}\right)\right]
\end{eqnarray}
The coefficient $c_{10}(\mu)=c_{10}(m_W)$, and $\xi = \alpha_s(\mu)/\alpha_s(m_W)$. 

The Wilson coefficients at the scale $m_W$ receive the following contributions from $W$ loops:
\begin{eqnarray}
c_7(m_W)&=&-\frac{1}{2}A(x),\;\;c_8(m_W)=-\frac{1}{2}E(x),\;\;c_{10}(m_W)=\frac{1}{s^2_W}[B(x)-C(x)],
\nonumber \\
c_9(m_W)&=&-c_{10}(m_W)-4C(x)-D(x)+\frac{4}{9}.
\end{eqnarray}
The functions $A(x)$,... are:
\begin{eqnarray}
A(x)&=& \frac{x}{z^3}\left(\frac{2}{3}x^2+\frac{5}{12}x-\frac{7}{12}-(\frac{2}{3}x^2-x)\frac{\ln x}{z}\right),
\\
B(x)&=& \frac{x}{4z} (-1+\frac{\ln x}{z}),
\nonumber \\
C(x)&=&\frac{x}{4z}\left(\frac{1}{2}x-3+(\frac{3}{2}x+1)\right)\frac{\ln x}{z},
\nonumber \\
D(x)&=&\left(-\frac{19}{36}x^3+\frac{25}{36}x^2+(-\frac{1}{6}x^4+\frac{5}{3}x^3-3x^2+\frac{16}{9}x-\frac{4}{9})\frac{\ln x}{z}\right),
\nonumber \\
E(x)&=&\frac{x}{2z^3}\left(\frac{1}{2}x^2-\frac{5}{2}x-1+3x\frac{\ln x}{z}\right),
\nonumber
\end{eqnarray}
where $x=(m_t/m_W)^2$, and $z=x-1$.
The values of the Wilson coefficients are calculated at the scale $\mu \simeq m_b$,
for $m_b=4.8$ GeV, $\Lambda_{\bar{\rm MS}}=250$ MeV and $m_{\rm top} = 174$ GeV.
The other four coefficients turns out to be very small, i.e. at this scale they receive the following values: 
$c_3=0.017$, $c_4=-0.037$, $c_5=0.010$, and $c_6=-0.046$.

\subsubsection{{\it Inclusive radiative and semileptonic decays}}

To avoid the uncertainty in $m_b$, it is customary to express the branching ratios 
$BR(b \to s \gamma)$ and $BR(b\to s \ell^+\ell^-)$ in terms
of the dominant semileptonic branching ratios $BR(b\to c{\ell}{\bar\nu}_{\ell})$:
\begin{eqnarray}
R_{\gamma}&\equiv&\frac{\Gamma (b \to s \gamma)}{\Gamma (b\to c{\ell}{\bar\nu}_{\ell})}=
\frac{6\alpha_{\rm em}}{\pi \lambda g(m_c^2/m_b^2)}\;\frac{|V^*_{ts}V_{tb}^{}|^2}{|V_{cb}|^2}\;|c_7(m_b)|^2,
\\
R_{\ell^+\ell^-}&\equiv&\frac{\Gamma (b\to s \ell^+\ell^-)}{\Gamma (b\to c{\ell}{\bar\nu}_{\ell})}
 \\
&=&\left(\frac{\alpha_{\rm em}}{4\pi}\right)^2 \frac{2|V_{cs}|^2}{\lambda g(m_c^2/m_b^2)} \;
\left[F_1(|c_9|^2+|c_{10}|^2)+F_3c_7c_9+F_2|c_7|^2\right],
\nonumber
\end{eqnarray}
where the phase space factor $g(m_c^2/m_b^2)$ and the QCD correction factor $\lambda$ for the semileptonic 
process are well known \cite{dt4,barg}. We have used $g(m_c^2/m_b^2)=0.507$ and $\lambda=0.888$. 
The phase space integration from $(q^2)_{\rm min}= (2m_{\ell}/m_b)^2$ to $(q^2)_{\rm max}= (1-m_s/m_b)^2$ give
the following values for constants $F_i$ \cite{dt2}:
\begin{eqnarray}
F_1=1,\;\; F_3=8,\;\; for\;\; (q^2)_{\rm min}\cong 0,\;\;(q^2)_{\rm max} \cong 1,
\nonumber\\
F_2=32\ln(m_b/2m_{\ell}),\;\; for\;\; (q^2)_{\rm max} \cong 1,\;\;\ell=e,\mu,\tau.
\end{eqnarray}
The SM theoretical prediction for the inclusive radiative decay, up to NLO \cite{ciuc} 
in $\alpha_s \ln(m_w/m_b)$,
\begin{eqnarray}
BR(B \to X_s \gamma )_{\rm NLO}=(3.30\pm 0.32)\times 10^{-4},
\end{eqnarray}
is considerably larger than the lowest-order result\cite{gr}:
\begin{eqnarray}
BR(B \to X_s \gamma )_{\rm LO}=(2.46\pm 0.72)\times 10^{-4}.
\end{eqnarray}
Gambino and Misiak \cite{gam} performed a new analysis and reported a higher
short-distance (SD) result:
\begin{eqnarray}
BR(B \to X_s \gamma )_{\rm SD}=(3.73\pm 0.30)\times 10^{-4}.
\end{eqnarray}

Let us now present and discuss the experimental results:

Based on $9.7\times 10^6$ analysed $B{\bar B}$ pairs from $\Upsilon\,(4s)$, the  
CLEO Collaboration reported two years ago the following inclusive branching ratio \cite{cleo}:
\begin{equation}
BR(B\to X_s\gamma) = (3.22 \pm 0.40) \times 10^{-4}.
\end{equation}

Analysing $33\times 10^6 \;B{\bar B}$ pairs, a BaBar reported  20\% larger rate \cite{gambab}:
\begin{equation}
BR(B\to X_s\gamma) = 3.88  \times 10^{-4},
\end{equation}
and they also published a measurement of the inclusive branching ratio, obtained by summing up exclusive modes,
which is even larger than the first one \cite{gambab1}:  
\begin{equation}
BR(B\to X_s\gamma) = \sum_i BR(B\to K^*_i\gamma) = 4.3 \times 10^{-4}.
\end{equation}
A few years ago I was reporting that the inclusive branching ratio will increase with the number of 
events analysed, up to the certain point, of course \cite{josip}.

The Belle Collaboration reported the first results on inclusive semileptonic decay \cite{Belle}:
\begin{equation}
BR(B\to X_s\ell^+\ell^-) = \left(6.1\pm 1.4{+1.4 \atop -1.1}\right) \times 10^{-6},
\end{equation}
which is in fair agreement with our theoretical predictions for the $m_{\rm top}\cong 180$ GeV \cite{dt1,dt4,dt2}. 
See for example Fig.1, in Ref's \cite{dt1,dt2}. Note here that we estimated 
the $e$--$\mu$ rate for the inclusive \cite{dt1,dt2}
\begin{eqnarray}
BR(b\to s e^+e^-)/BR(b\to s\mu^+\mu^-) \simeq 1.4\, -\,1.6.
\end{eqnarray}
and found that the $e$--$\mu$ ratio has a weak dependence of $m_{\rm top}$.

\subsubsection{\it Exclusive radiative and semileptonic decays}

Exclusive modes are, in principle, affected by large theoretical uncertainties
due to the poor knowledge of non-perturbative dynamics and of a correct 
treatment of large recoil-momenta, which determine the form factors.

First we have to define the hadronic form factors.
The Lorentz decomposition of the penguin matrix elements for $(q=p-k)$ is:
\begin{eqnarray}
&& \langle K^*(k)|{\bar s}\sigma_{\mu\nu}q^{\nu}(1+\gamma_5)b|B(p)\rangle
=i\varepsilon_{\mu\nu\rho\tau}\epsilon^{*\nu}(q)(p+k)^{\rho}q^{\tau}T_1(q^2)
\nonumber\\
&&+T_2(q^2)\left[\epsilon^{*}_{\mu}(q)\left(m^2_B-m^2_{K^*}\right)- (p\epsilon^*(q))(p+k)_{\mu}\right]
\nonumber\\
&&+T_3(q^2)(p\epsilon^*(q))\left[q_{\mu}-\frac{q^2}{m^2_B-m^2_{K^*}}(p+k)_{\mu}\right],
\end{eqnarray}
with $T_1(0)=T_2(0)$ as a consequence of the spin symmetry. Note that the last term in the square bracket
vanishes for real photon. Similarly, for semileptonic (and/or non-leptonic) decays, we have
\begin{eqnarray}
&& \langle K^*(k)|{\bar s}\gamma_{\mu}(1-\gamma_5)b|B(p)\rangle 
=-i\varepsilon_{\mu\nu\rho\tau}\epsilon^{*\nu}(k)(p+k)^{\rho}q^{\tau}V
\\
&&+\epsilon^{*}_{\mu}(k)\left(m^2_B-m^2_{K^*}\right)A_1- (q\epsilon^*(k))(p+k)_{\mu}A_2
\nonumber\\
&&+\left(q\epsilon^*(k)\right)(m_B+m_{K^*})\left(q_{\mu}/q^2\right)
\left[2m_{K^*}A_0-(m_B-m_{K^*})(A_1-A_2)\right],
\nonumber
\end{eqnarray}
with the corresponding definitions of the relevant form factors
\begin{eqnarray}
V=\frac{V(q^2)}{(m_B+m_{K^*})},\; V(q^2)=\frac{V(0)}{1-\frac{q^2}{m^2_{1^-}}},
A_0=\frac{A_0(q^2)}{(m_B+m_{K^*})}, A_0(q^2)=\frac{A_0(0)}{1-\frac{q^2}{m^2_{0^-}}},
\nonumber \\
A_1=\frac{A_1(q^2)}{(m_B-m_{K^*})}, A_1(q^2)=\frac{A_1(0)}{1-\frac{q^2}{m^2_{1^+}}},
A_2=\frac{A_2(q^2)}{(m_B+m_{K^*})}, A_2(q^2)=\frac{A_2(0)}{1-\frac{q^2}{m^2_{1^+}}},
\nonumber
\end{eqnarray}
\begin{eqnarray}
A_0=\frac{m_B+m_{K^*}}{2m_{K^*}}A_1(0)-\frac{m_B-m_{K^*}}{2m_{K^*}}A_2(0).
\end{eqnarray}
In the above $V$ and $A_i,\;(i=0,1,2)$ form factors, the $q^2$ for semileptonic is determined by the 
invariant lepton pair mass squared, while for the two-body non-leptonic decays (calculated in a
factorization approximation) it is the mass squared of the factorized meson.

Finally, the operator $O_7$, taking into account the gauge condition, the current
conservation, the spin symmetry, and for real photon, gives a following
hadronization rate $R_{K^*}$ \cite{dlt,desh}:
\begin{eqnarray}
R_{K^*}=
\frac{{\Gamma(B\to K^* \gamma)}}{{ \Gamma(b\to s\gamma)}} 
 = \left[\frac{m_b(m^2_B-m^2_{K^*})}{m_B(m^2_b-m^2_s)}\right]^3
\left(1+\frac{m^2_s}{m^2_b}\right)^{-1} |T_1^{K^*}(0)|^2. 
\end{eqnarray}
In Table 1 we give a few typical results for the hadronic form factors, while in Table 2
the typical hadronization rates are given for different types of the form factor estimates. 
\renewcommand{\arraystretch}{1.8}
\begin{table}
\caption{Comparison of a few different reults on form factors at $q^2=0$.}
\begin{center}
\begin{tabular}{|c|c|c|c|c|c|}
\hline 
 $ \rm{Form \atop factors}   $  & $\rm{Ref.\cite{col} \atop (3pt SR)}$ & $ \rm{Ref.\cite{ali} \atop (LCSR)}$ &  
 $ \rm{Ref.\cite{aliev} \atop (LCSR)} $ &  $ \rm{Ref.\cite{del} \atop (lattice+LCSR)} $ &  
 $ \rm{Ref.\cite{ball} \atop (LCSR)} $ \\
\hline \hline
 $ V^{K*}(0) $  & $ 0.47\pm0.3 $  & $  0.38\pm0.08  $ & 
 $  0.45\pm0.08 $ & $ - $  & $ 0.46\pm0.07 $   \\         
\hline
 $ A_1^{K*}(0) $  & $  0.37\pm0.03 $ & $ 0.32\pm0.06 $    & 
 $ 0.36\pm0.05 $  & $ 0.29{+0.4 \atop -0.03} $ & $ 0.34\pm0.05 $    \\                       
\hline
 $ A_2^{K*}(0) $  & $ 0.40\pm0.03 $ & $ - $ &  $ 0.40\pm0.05 $ & $ - $ & $ 0.28\pm0.04 $  \\
\hline 
 $ T_1^{K*}(0) $  & $  0.38\pm0.06 $  & $ 0.32\pm0.05 $  & 
 $ 0.34\pm0.10 $ & $ 0.32{+0.04 \atop -0.02} $ & $ 0.38\pm0.06 $   \\         
\hline
 $ T_3^{K*}(0) $  & $  0.6 $ & $ - $  & $ 0.26\pm0.10 $ & $ - $   & $ 0.26\pm0.04 $    \\                       
\hline
 $ V^{\rho}(0) $  & $  0.6\pm0.2  $ & $ 0.35\pm0.07  $  & 
 $ 0.37\pm0.07 $ &  $ 0.35{+0.06 \atop -0.05} $ &  $ 0.34\pm0.05  $  \\
\hline 
 $ A_1^{\rho}(0) $  & $ 0.5\pm0.1 $  & $ 0.27\pm0.05 $  & 
 $  0.30\pm0.05 $ & $ 0.27{+0.05 \atop -0.04} $  & $ 0.26\pm0.04 $   \\         
\hline
 $ A_2^{\rho}(0) $  & $  0.4\pm0.2  $ & $ 0.28\pm0.05 $   & 
 $ 0.33\pm0.05$ & $ 0.26{+0.05 \atop -0.03} $  & $ 0.22\pm0.03 $    \\                       
\hline
 $ T_1^{\rho}(0) $  & $ - $ & $ 0.24\pm0.07 $  &  $  0.30\pm0.10  $ &  $ 0.32\pm0.06 $ &  $ 0.29\pm0.04 $  \\
\hline 
$ T_3^{\rho}(0) $  & $  - $ & $ - $  &  $ 0.20\pm0.10 $ &  $ - $ &  $ 0.20\pm0.03 $  \\
\hline 
\end{tabular}\\
\label{t:tab1}
\end{center}
\end{table}

\renewcommand{\arraystretch}{1.6}
\begin{table}
\caption{Comparison of the results for the hadronization rate $R_{K^*}$[\%]}
\begin{center}
\begin{tabular}{|c|c|c|c|}
\hline 
$ \rm {Authors} $ & $ \rm{Reference} $ & $ R_{K^*} $[\%] & $ \rm{Model} $ \\
\hline \hline
$\rm O'Donnel \;(1986) $ & $ \cite{odon} $ & $ 97.0 $ & $ - $  \\
\hline
$\rm Deshpande \;et \;al. \;(1987) $ & $ \cite{des} $ & $ 7 $ & $\rm CQM $  \\
\hline
$\rm Deshpande \;et \;al. \;(1988) $ & $ \cite{dlt} $ & $ 6 $ & $\rm RCQM $  \\
\hline
$\rm Altomari \;(1988) $ & $ \cite{alto} $ & $ 4.5 $ & $\rm CQM $  \\
\hline
$\rm Deshpande, \;Trampeti$\' c$ \;(1989) $ & $ \cite{desh} $ & $ 6-14 $ & $ \rm RCQM $  \\
\hline
$\rm Ali, \;Mannel \;(1991) $ & $ \cite{ali} $ & $ 28-40 $ & $ \rm QCD SR $  \\
\hline
$\rm Faustov,\; Galkin \;(1992) $ & $ \cite{faus} $ & $ 6.5 $ & $ \rm RCQM $  \\
\hline
$\rm Colangelo \;et \;al. \;(1993) $ & $ \cite{col} $ & $ 16\pm3 $ & $ \rm 3pt SR $  \\
\hline
$\rm Casalbuoni \;et \;al. \;(1993) $ & $ \cite{casa} $ & $ 8 $ & $ \rm{{CSL+SS HQET} \atop {+Experiment}} $  \\
\hline
$\rm Atwood, \;Soni \;(1994) $ & $ \cite{soni} $ & $ 1.6-2.5 $ & $ RCQM $  \\
\hline
$\rm Bowler \;et \;al.\;(1994) $ & $ \cite{bow} $ & $ 9.0\pm3.0\pm1.0 $ & $ \rm QCD \;on \;lattice $  \\
\hline
$\rm Ali, \;Sima\; (1994) $ & $ \cite{alis} $ & $ 12\pm2 $ & $ \rm QCD SR $  \\
\hline
$\rm Bernard \;et \;al.\; (1994) $ & $ \cite{bern} $ & $ 6\pm1.2\pm3.4 $ & $ \rm QCD\; on \;lattice $  \\
\hline
$\rm Burford \;et \;al. \;(1995) $ & $ \cite{burf} $ & $ 15.0-35.0 $ & $ \rm QCD\;on \;lattice $  \\
\hline
$\rm Veseli, \;Olsson\;  \;(1996) $ & $ \cite{ves} $ & $ 16.8\pm6.4 $ & $ \rm HQET $  \\
\hline
$\rm Aliev \;(1997) $ & $ \cite{aliev} $ & $ 13\pm4 $ & $ \rm LCSR $  \\
\hline
$\rm Ball,\; Braun \;(1998) $ & $ \cite{ball} $ & $ 16\pm3 $ & $ \rm LCSR $  \\
\hline
$\rm Del Debbio \;et \; al. \;(1998) $ & $ \cite{del} $ & $ 12{+2 \atop -1} $ & $ \rm{{LCSR+lattice} \atop {+constraints}} $  \\
\hline
$\rm Mohanta \;et \;al. \;(1999) $ & $ \cite{moha} $ & $ 12 $ & $ \rm COQM $  \\
\hline
$\rm Asatryan \;et \;al.\; (1999) $ & $ \cite{asa} $ & $ 16 $ & $ \rm{{NLL \;for \;} \atop {B\to K^*\gamma}} $  \\
\hline
$\rm Bosch, \;Buchalla \;(2002) $ & $ \cite{buch} $ & $ 22 $ & $ \rm{{NLO\; pQCD \;type} \atop {exclusive/inclusive}} $  \\
\hline
\end{tabular}\\
\end{center}
\label{t:tab2}
\end{table}
Since the first calculation of the hadronization rate
$R_{K^*} \approx 7$\%  by Deshpande at al. \cite{des},
a large number of papers have reported $R_{K^*}$
from the range of $3$ to an unrealistic $90$\%.
Different methods have been employed, from
quark models \cite{des,dlt,desh,alto,moha}, QCD sum rules \cite{col,ali}, HQET and chiral symmetry \cite{casa}, 
QCD on the lattice \cite{bow}, 
light cone sum rules \cite{ball}, to the
perturbative QCD type of evaluations of exclusive modes \cite{asa,buch}.

Concerning Ref. \cite{buch} we have to comment that even in such very complex evaluations
of exclusive modes, the hadronic form factor $T_1(0)$ was not included as a part of 
first principal pQCD calculations, but was rather used as an input from other sources \cite{ball}.
Clearly, the final results of Ref \cite{buch} crucially depend on the authors choice of $T_1(0)$.

In any event, the above form factor will be obtained
in the future from first principle calculations
on the lattice. 
Recently, seems that the hadronization rate $R_{K^*}$ in radiative decay calculations
has stabilized around 10\%.
\\

Exclusive semileptonic B decay rates, estimated for $m_{\rm top}\simeq 180$ GeV in Refs.\cite{dt1,dt4,dt2},
\begin{eqnarray}
BR(B\to K e^+e^-)/BR(b\to s e^+e^-) \cong 0.08,
\nonumber\\
BR(B\to K^* e^+e^-)/BR(b\to s e^+e^-) \cong 0.20,\\
BR(B\to K^* e^+e^-)/BR(B\to K^*\mu^+\mu^-)\cong 1.23,
\end{eqnarray}
were later confirmed by other authors. 
\\

The first measurements by the Belle Collaboration \cite{belle} 
\begin{eqnarray}
&&BR(B \to K \ell^+\ell^-) = (0.75 {+0.25 \atop -0.21} \pm 0.19) )\times 10^{-6},
\nonumber\\
&&BR(B \to K \mu^+\mu^- ) = (0.99 {+0.40 \atop -0.32} {+0.13 \atop -0.14})\times 10^{-6},
\end{eqnarray}
are in good agreement with theory.
\\

Concerning the rare B decay to the orbitally excited strange mesons, the first CLEO \cite{cl1} observation has recently been 
confirmed by Belle \cite{belle1,belle2}. These important experimental measurements provide a crucial challenge to the
theory. 
The exclusive radiative B decays into higher spin-1 resonances are described by a formula similar to above:
\begin{eqnarray}
R_{K^{**}}=
\frac{{\Gamma(B\to K^{**} \gamma)}}{{ \Gamma(b\to s\gamma)}} 
 = [\frac{m_b(m^2_B-m^2_{K^{**}})}{m_B(m^2_b-m^2_s)}]^3
(1+\frac{m^2_s}{m^2_b})^{-1} |T_1^{K^{**}}(0)|^2. 
\end{eqnarray} 
The $K^{**}$ represent all the higher resonances. 
Most of these theoretical approaches rely on non-relativistic quark models \cite{dlt,ali,alto}, HQET \cite{ves}, relativistic
model \cite{ebert}, and LCSR \cite{safir}. 
Different results for the hadronization rate
$R_{K^{**}}$ are presented in Table 3.
Note, however, that there is a large spread between different results, because of their different 
treatments of long-distance effects.
\renewcommand{\arraystretch}{1.6}
\begin{table}
\caption{Comparison of few a different results for the rate $R_{K^{**}}$ [\%]}.
\begin{center}
\begin{tabular}{|c|c c c c c c|}
\hline 
 $  $  &     &      & $  R_{K^{**}} $[\%] &     &      &    \\         
\hline 
 $ \rm Meson   $  & $\rm Ref.\cite{dlt} $ & $\rm Ref.\cite{alto}$ & $\rm Ref.\cite{ali} $ &  $\rm Ref.\cite{ves} $ &  
 $\rm Ref.\cite{ebert} $ & $\rm Ref.\cite{safir}$ \\
\hline \hline
 $ K(494) $  &     &      & $ \rm forbidden $ &     &      &    \\         
\hline 
 $ K^*(892) $  & $ 6 $  & $  4.5  $ & $ \rm 3.5-12.2 $ & $ 16.8\pm6.4 $  & $ 15\pm3 $   & $ 10.0\pm 4.0 $ \\         
\hline 
 $ K^*(1430) $  &   &     & $ \rm forbidden $ &   &    & \\         
\hline 
 $ K_1(1270) $  & $\rm forb. $  & $ \rm forb./6.0  $ & $ \rm 4.5-10.1 $ & $ 4.3\pm1.6 $  & $ 1.5\pm0.5 $  & $ 2.0\pm0.8 $ \\         
\hline 
 $  K_1(1400)  $  & $ 7 $  & $ \rm forb./6.0 $ & $ \rm 6.0-13.0 $ & $ 2.1\pm0.9 $  & $ 2.6\pm0.6 $  & $ 0.9\pm0.4 $ \\         
\hline 
 $ K^*_2(1430) $  & $  $  & $ 6.0 $ & $ \rm 17.3-37.1 $ & $ 6.2\pm2.0 $  & $ 5.7\pm1.2 $   & $ 5.0\pm2.0 $ \\         
\hline 
 $ K^*(1680) $  & $  $  & $ 0.9 $ & $ \rm 1-1.5 $ & $ 0.5\pm0.2 $  & $  $  & $ 0.7\pm0.3 $ \\         
\hline 
 $ K_2(1580) $  & $  $  & $ 4.4 $ & $ \rm 4.5-6.4 $ & $ 1.7\pm0.4 $  & $  $  & $  $ \\         
\hline 
$ K(1460) $  &  &  & $ \rm forbidden $ &   &   &  \\         
\hline 
 $ K^*(1410) $  & $  $  & $ 7.3 $ & $ \rm 7.2-10.6 $ & $ 4.1\pm0.6 $  & $  $   & $ 0.8\pm0.4 $ \\         
\hline 
 $ K^*_0(1950) $  &   &  & $ \rm forbidden $ &   &    &  \\         
\hline 
 $ K_1(1650) $  & $  $  & $\rm not \;given $ & $\rm not \;given $ & $ 1.7\pm0.6 $  & $  $   & $ 0.8\pm0.3 $ \\         
\hline 
\end{tabular}\\
\label{t:tab3}
\end{center}
\end{table}\\

The modes based on $b\to d \gamma$ represent a powerful way of determining the CKM ratio $|V_{td}/V_{ts}|$.
If long-distance and other non-perturbative effects
are neglected, two exclusive modes
are connected by a simple relation \cite{tram}:
\begin{equation}
BR(B \rightarrow \rho \gamma) =
\xi^{2} \left| V_{td}/V_{ts} \right|^{2}
BR(B \rightarrow K^{*} \gamma),
\end{equation}
where $\xi$ measures the SU(3) breaking effects.
They are typically of the order of 30\% \cite{ball}.
Misiak has reported the following short-distance contributions to the branching ratios \cite{mis}: 
\begin{equation}
BR(b \to d \gamma)=1.61\times 10^{-5};\; 
BR(B^+ \rightarrow \rho^+ \gamma) =  (1-4) \times 10^{-6},
\end{equation}
\begin{equation}
BR(B^0 \rightarrow \rho^0 \gamma) =
BR(B^0 \rightarrow \omega \gamma) =
(0.5-2) \times 10^{-6}.
\end{equation}
The simple isospin relations are valid for the above decay modes:
\begin{eqnarray}
\Gamma(B^+ \rightarrow \rho^+ \gamma)=2\Gamma(B^0 \rightarrow \rho^0 \gamma)=2\Gamma(B^0 \rightarrow \omega \gamma).
\end{eqnarray}

This year, experimental results for exclusive radiative and semileptonic decay modes, based on $33\times 10^6 \;B{\bar B}$ pairs, 
are coming from the Belle Collaboration \cite{belle2}
\begin{eqnarray}
&&BR(B^0 \to K^{*0} \gamma) = \left(4.08 {+0.35 \atop -0.33} \pm 0.26\right) \times 10^{-5},
\nonumber\\
&&BR(B^+ \to K^{*+} \gamma) = \left(4.92 {+0.59 \atop -0.54} {+0.38 \atop -0.37}\right) \times 10^{-5},
\nonumber\\
&&BR(B \to K^{*}_2(1430) \gamma) = \left(1.50 {+0.58 \atop -0.53} {+0.11 \atop -0.13}\right) \times 10^{-5},
\end{eqnarray}
\begin{eqnarray}
{\cal A}_{CP} = \frac{\Gamma({\bar B} \to {\bar K}^*\gamma)-\Gamma( B \to K^*\gamma)}
{\Gamma({\bar B} \to {\bar K}^*\gamma)+\Gamma( B \to K^*\gamma)}=
\left(3.2 {+6.9 \atop -6.8} \pm 2.0\right)\% \cong 0.
\end{eqnarray}
Using the latest results for inclusive and exclusive branching ratios, we have obtained
the following central value for the so-called hadronization rate: $R_{K^*}^{exp} \simeq 10$\%,
which is in excellent agreement with the theory.

The BaBar Collaboration \cite{babar1} produced the latest experimental results on exclusive semileptonic B decays:
\begin{eqnarray}
&&BR(B \to K \ell^+\ell^-) = \left(0.78 {+0.24 \atop -0.20} \pm 0.26 {+0.11 \atop -0.18}\right)\times 10^{-6},
\nonumber\\
&&BR(B \to K^* \ell^+\ell^- ) = \left(1.68 {+0.68 \atop -0.58} \pm 0.26 \pm0.28\right)\times 10^{-6}.
\end{eqnarray}

Measurements for exclusive modes based on the quark $b \to d \gamma$ transition 
were recently reported by the BaBar Collaboration \cite{babar2}:
\begin{eqnarray}
&&BR(B^{0} \rightarrow \rho^{0} \gamma) < 1.5 \times 10^{-6}
\nonumber\\
&&BR(B^{+} \rightarrow \rho^{+} \gamma) < 2.8 \times 10^{-6}
\nonumber\\
&&BR(B \rightarrow \rho \gamma)/BR(B \rightarrow K^* \gamma)<0.34;
\end{eqnarray}
they are considerably lower than the first CLEO results \cite{cl}.
However, the isospin relations (32) are nicely satisfied.
From the BaBar measurements we obtain the following ratio of CKM:
\begin{eqnarray}
|V_{td}|/|V_{ts}|<0.64\, -\,0.76.
\end{eqnarray}

{\it Note about quark models}
\\
In principle there are two types of models for describing hadrons, i.e.
quark models: {\it non-relativistic potential} and {\it relativistic}
models. They are all represented mainly by the constituent quark model (CQM) and the MIT bag model.

Almost all quark models describe the static properties of ground state hadrons with $15$\% 
accuracy. In particular, the CQM and the MIT Bag model have been very useful when computing the mass spectrum 
and static properties such as charge radii, magnetic moments, $ (g_A/g_V)_{p,n}$, etc., of ground state baryons.
Apart from the fact that the MIT Bag model is essentially the solution of the Dirac equation with 
boundary conditions, we have to note that this model is {\it static}, which 
is certainly disadvantage.
The MIT Bag model also has problems in describing the particle's higher excited states.

On the other hand 
the non-relativistic CQM (harmonic oscillator type, etc.) \cite{stech,isgur}
could take into account 
the motion of the particle as a whole, but it is not well grounded conceptually.
However, these models, based on Gaussian wave functions, give us the possibility to
compute effects coming from the internal quark motions as well as from the moving particle as a whole.
These models have also been successful in computing mesonic pseudo-scalar, vector and tensor
form factors.\\

{\it Note about HQET}
\\
The physical essence of the {\it Heavy Quark Symmetry} lies in the fact that the internal dynamics of
the heavy hadrons becomes independent of heavy quark mass $m_Q$ and the quark spin when $m_Q$ is 
sufficiently heavy. The heavy quark becomes a static source of colour fields in its rest frame. 
The binding potential is flavour-independent and spin effects fall like $1/m_Q$.
Light quarks and gluons in the hadron are the same whether $Q=c$ or $Q=b$ (as $m_{c,b} \to \infty$).

In HQET the heavy quark moves with the hadron's velocity $v$, so that the heavy quark momentum is
\begin{eqnarray}
P^{\mu}_Q = m_Q v^{\mu}  + k^{\mu},
\end{eqnarray}
where $k^{\mu}$ represents the small residual momenta. 

The velocity $v_{\mu}$ in heavy quark rest frame,
according to the Georgi's covariant description, has the very simple form $v_{\mu}=(1,{\vec 0})$.
The heavy quark propagator has to be modified accordingly:
\begin{eqnarray}
\lim_{{m_Q} \rightarrow \infty} \frac{i}{{\not \!\!P_Q} - m_Q} 
= \frac{i}{v\cdot k}\frac{1+{\not \!v}}{2} + {\cal O}(k/m_Q)= \frac{i}{v\cdot k}\frac{1+{\not \!v}}{2} + ....
\end{eqnarray}
The residual momentum is in effect a measure of how off-shell the heavy quark is. The HQET is valid for 
$m_Q >> |k| \simeq \Lambda_{\rm QCD}$.

Applying the limit $m_Q \rightarrow \infty$ on the covariant form of QCD Lagrangian for heavy quarks,
we obtain:
\begin{eqnarray}
{\cal L}_{\rm QCD}= {\bar Q}(i{\not \!\!D} - m_Q)Q\;\;\longrightarrow
{\cal L}_{\rm HQET}= {\bar h}^{(Q)}_v i(v\cdot D)h^{(Q)}_v,
\end{eqnarray}
From the above Lagrangian ${\cal L}_{\rm HQET}$, we obtain the following Feynman rule for the quark--quark--gluon
vertex in HQET: $igT^av_{\mu}$.

It is very important to note here that $h^{(Q)}_v$ destroys a heavy quark of the 4-velocity $v^{\mu}$ and 
{\it $\underline{does \;not}$} create a correct antiquark.

Finally, this theory uses the mass of the heavy quark as an expansion parameter, yillding predictions 
in terms of powers of $1/m_Q$.

\subsection{{\it Long-distance and other small contributions}\\ {\it to inclusive and exclusive B decays}}

{\it Long-distance corrections}
\\
First note that long-distance contributions for exclusive decays could not be computed
from first principles without the knowledge of the hadronization process. However, it is possible to
estimate them phenomenologically \cite{pak}. 

The operators $O_{1,2}$ contain the $\bar cc$ current. So
one could imagine the $\bar cc$ pair propagating through a long distance,
forming intermediate $\bar cc$ states  
(off-shell $J/\psi$'s), which turn into a photon via the vector meson dominance (VMD) mechanism.
Application of the VMD mechanism on the quark level was used by Deshpande et al. \cite{dht}.
Such an approach, with a careful treatment of the decay amplitude by the Lorentz and electromagnetic gauge invariance,
i.e. by cancelling the contributions coming from longitudinal photons, makes it possible 
to form the total (short- plus long-distance) amplitude for the $b \to d(s) \gamma$ decay \cite{dht}
\begin{eqnarray}
&& M(b\rightarrow d \gamma)|_{total} 
= -\frac{eG_F}{2\sqrt{2}}\left[V_{td}^{*}V_{tb}^{}\left(\frac{m_b}{4\pi^2}c_7(m_b)
-\frac{2}{3}a_2\sum_i \frac{g_{\psi_i}^2(0)}{m_{\psi_i}^2 m_b}\right)\right.\\
&&
\left.-\frac{a_2}{m_b}V_{ud}^{*}V_{ub}^{} \left(\frac{2}{3}\sum_i
\frac{g_{\psi_i}^2(0)}{m_{\psi_i}^2}-\frac{1}{2}\frac{g_\rho^2(0)}{m_\rho^2}-{1\over 6}
\frac{g_\omega^2(0)}{m_{\omega}^2}\right)\right]{\bar d}\sigma^{\mu\nu}(1+\gamma_5)b F_{\mu\nu}.
\nonumber
\end{eqnarray}
If in the above equation we replace the $d$ by the $s$ quark
and forget the last three terms, then
we obtain the total amplitude for the $b \to s \gamma$ decay.
It is important to note that we have found strong suppression when
extrapolating $g_{\psi}(m_{\psi}^2)$ to $g_{\psi}(0)$: 
$g^2_{\psi(1S)}(0)/g^2_{\psi(1S)}(m^2_{\psi}) = 0.13 \pm 0.04$ \cite{dht}. This fact has to be taken into account
in any other approach (LCSR, pQCD, lattice-QCD, etc.) to the long-distance problem \cite{gp}.

The long distance contributions to an inclusive amplitude and to its exclusive mode
are all found to be small, typically of one order of magnitude below the short distances \cite{dht}.\\

{\it Other small corrections}
\\
Other small corrections to the $b \to d(s) \gamma$ transitions come from spectator quark
contributions \cite{don},
non-perturbative effects \cite{vol} and from the fermionic and bosonic loop effects \cite{marc}.
\\
(i) Donoghue and Petrov found that the spectator contributions to rare inclusive B decays are about $5$\%, i.e.
they give the following rise to the branching ratio \cite{don}:
\begin{equation}
\Delta \left(BR(B\to X_s\gamma)/BR(b\to s\gamma)\right) \simeq +1.05;
\end{equation}
\\
(ii) Non-perturbative corrections up to the $\Lambda^2/m^2_c$ order were estimated  
by Voloshin \cite{vol}.
They gives the following rise to the branching ratio:
\begin{equation}
\Delta \left(BR(b\to s\gamma)\right) \simeq +3\%;
\end{equation}
\\
(iii) Czarnecki and Marciano calculated the leading electroweak corrections via 
fermionic and bosonic loops. In particular, the vacuum polarization renormalization of $\alpha$ by the fermionic loops,
contributions from quarks and leptons in the W propagator loops, the two-loop diagrams where a virtual photon exchange 
gives a short-distance logarithmic contribution, etc. These corrections reduce $BR(b\to s\gamma)$ by $\sim 8$\% 
\cite{marc}, i.e.
\begin{equation}
\Delta \left(BR(b\to s\gamma)/BR(b\to c e{\bar\nu})\right) \simeq -(8\pm2)\%.
\end{equation}
Note that $\alpha_{\rm em}=1/137$ for a real photon was used.
 
However, all above corrections never exceed an overall $\sim 10$\%, and on top of that there is a 
cancellation among them!
So it turns out that the inclusive branching ratio is stable and agrees well with measurements.

\subsection{\it Non-leptonic B decays}

Non-leptonic processes at the quark level involve gluons and ${\bar q}q$ pairs, i.e. they are dominated by transitions
$b \to s(d)g$ and $b \to s(d){\bar q}q$ \cite{guba,hou}. The following non-leptonic B meson decay properties are very
important:\\
(i)   they play major a role in the determination of the unitarity triangle parameters: $\alpha$, $\beta$, and $\gamma$;\\
(ii)  there are three decay classes:

1. pure 'tree' contributions,

2. pure 'penguin' contributions,

3. 'tree + penguin' contributions;\\
(iii) there are two penguin topologies:

1. gluonic (QCD) penguins,

2. electroweak (EW) penguins;\\
(iv)  the photon in EW penguin could be real ($\gamma$) or virtual ($\gamma^*\to {\bar q}q, {\bar {\ell}}\ell$);\\
(v)   there are two types of decay modes:

1. the $b\to s {\bar q}q$ mode,
 
2. the $b\to d {\bar q}q$ mode. 

The experimental signatures for such charmless 
transitions are exclusive decays such as $B \to \pi\pi,\; \pi K$, etc. For the $b \to s(d){\bar c}c$ 
transitions involving charm, the exclusive decays are the very well known $B \to
J/\psi K$,... and the less known  $B \to \eta_c K$,...
modes. These modes in general are not considerd to belong to the rare decays. However, the modes based on
$b \to s\;\eta_c$ are just an order of magnitude larger than the rare sector. So they are interesting enough to be discussed
in one of the next subsections.

\subsubsection{\it The $b\to s {\bar q}q$ and $b\to d {\bar q}q$ decay modes}

The calculations for these processes involve matrix elements of 
four-quark operators of dimension 6, and there are difficulties to estimate
these elements. An added complication here is that 
charmless hadronic decays can also arise through the tree Hamiltonian with $b \to u$ transition. A careful study of these
modes reveals that where penguins clearly dominate in  some, while the tree
contribution can be significant in others.

The calculation proceeds in two steps \cite{dt3}. First we obtain the effective
short-distance interaction including one-loop
gluon-mediated diagram (I). We then use the factorization approximation to derive the hadronic matrix elements 
by saturating with vacuum state in all possible ways (II). The resulting matrix elements 
involve quark bilinears between one meson state
and the vacuum, and between two meson states. These are estimated using 
relativistic quark model wave functions, 
lattice model calculations, light cone sum rules, the perturbative QCD type of approach, etc.

(I) To get a better understanding of the complete QCD-corrected weak Hamiltonian density we shall discuss the 
gluon-mediated penguin contribution. Dictated by gauge invariance, 
the effective FCNC $J_{\mu}$
contains, as in the electromagnetic case, two terms. The first, which is proportional to $G_1$, we call the charge radius, while
the second, proportional to $G_2$, is called dipole moment operator
\begin{eqnarray}
J_{\mu}= {\bar s}^i\frac{1}{2}\lambda_{ij}\left\{G_1 (\gamma_{\mu}q^2-q_{\mu}{\not \!q})L +
iG_2 \sigma_{\mu\nu}q^{\nu}( m_sL + m_bR)\right\}b^j .
\end{eqnarray}
Using the standard parametrization of the CKM matrix in the case of three doublets, $G_1$ is given by
\begin{eqnarray}
G_1= \frac{G_F}{\sqrt 2}\frac{g_s}{4\pi^2}{\sum_{k=u,c,t}}A_k F_1(x_k),\; A_k=V^*_{ks}V^{}_{kb},\; x_i=m^2_i/m^2_W,\;z_k=1-x_k,
\end{eqnarray}
where the modified Inami--Lim function $F_1(x_i)$ derived from the penguin
(1-loop) diagrams is \cite{lim}
\begin{eqnarray}
F_1(x_i)= \frac{x_k}{12}\left(\frac{1}{z_k}+\frac{13}{z^2_k}-\frac{6}{z_k^3}\right)+
\left[ \frac{2}{3z_k}-\frac{x_k}{6}\left(\frac{4}{z^2_k}+\frac{4}{z^3_k}-\frac{3}{z^4_k}\right)\right]\ln x_k.
\end{eqnarray}

Note that when the gluon is on-shell (i.e. $q^2=0$), the $G_1$ term vanishes. 
In the $q^2 {\not =}0$ cases both terms participate.
For a gluon exchange diagram (i.e. for the processes $b \to s(d)\;{\bar q}q$ where momentum transfer $q^2 {\not =}0$) 
we find that the $G_1$ contribution dominates over $G_2$,
and we can neglect $G_2$. 
At larger $q^2$, $G_1$ develops a small 
imaginary part, which is important for a discussion of CP violation.

\begin{figure}
 \resizebox{0.9\textwidth}{!}{%
  \includegraphics{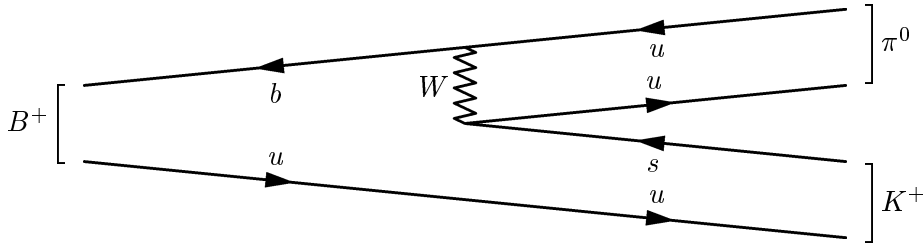}}
 \caption{Tree-dominated decay $B^+\to K^+\pi^0$.}
 \label{Zfig2}
\end{figure}

\begin{figure}
 \resizebox{0.9\textwidth}{!}{%
  \includegraphics{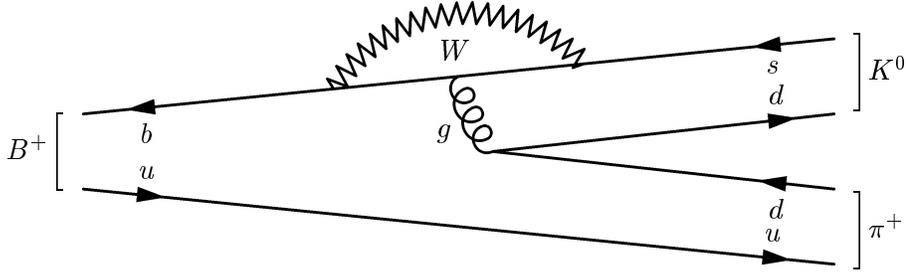}}
 \caption{Penguin-dominated decay $B^+\to K^0\pi^+$.}
 \label{Zfig3}
\end{figure}

Charmless decays also arise from the standard tree level interactions with the $b\to u$ transition. 
The effects of the tree level interaction could be
large in general. The most typical example are the two decay modes of the B$^+$ meson. The decay
$B^+ \to K^+\pi^0$ is dominated by the tree diagram, while the $B^+ \to K^0\pi^+$ is 
dominated by the penguin diagram; Fig.\ref{Zfig2}, and \ref{Zfig3}.

(II) The factorization approximation is in order.\\
From experience we know that non-leptonic decays are extremely difficult to handle. For example, the 
${\Delta I}=1/2$ rule in $K\to \pi\pi$ decays has not yet been fully understood.
A hughe theoretical
machinery has been applied to $K\to \pi\pi$ decays, producing only partial agreement with experiment \cite{mtt}. For 
energetic decays of heavy mesons (D,B), the situation is somewhat simpler. 
For these decays, the direct generation of a final meson by
quark current is indeed a good approximation.\\
According to the current-field identities, the currents are proportional to interpolating stable or quasi-stable hadron fields. The 
approximation now consists only in taking the asymptotic part of the full hadron
field, i.e. its ``in'' or ``out'' field.
Than the weak amplitude factorizes and is fully determined by the matrix elements of another current between the two remaining hadron states.
For that reason, we call this approximation the factorization approximation. Note that in replacing the interacting fields by 
the asymptotic fields, we have neglected any initial- or final-state interaction
of the corresponding particles. 
For B decays, this can be justified by the very simple energy argument that one very heavy object decays into two light 
but very energetic objects
whose interactions might be safely neglected. Also, diagrams in which a quark pair is created from vacuum will have small amplitudes
because these quarks have to combine with fast quarks to form the final-state meson. Note also that the $1/N_c$ expansion
argument provides a theoretical justification \cite{thoft} for the factorization approximation, 
since it follows the leading order in the $1/N_c$ expansion \cite{tt}. Here $N_c$ is the number of colours.\\
Each of the B-decay two-body modes might receive three different contributions.
As an example, we give one amplitude obtained from the effective 
weak Hamiltonian:
\begin{eqnarray}
A(B^+\to \pi^+ \pi^0)&=& L(\pi^0) \langle \pi^+|{\bar b}\gamma_{\mu}(1-\gamma_5)d|B^+\rangle 
\langle \pi^0|{\bar u}\gamma^{\mu}\gamma_5 u|0\rangle 
\\
&+&L(\pi^+) \langle \pi^+|{\bar u}\gamma_{\mu}\gamma_5 d|0\rangle \langle \pi^0|{\bar b}\gamma^{\mu}(1-\gamma_5)u|B^+\rangle
\nonumber\\
&+&L(B^+) \langle \pi^+\pi^0|{\bar u}\gamma_{\mu}(1-\gamma_5)d|0\rangle \langle 0|{\bar b}\gamma^{\mu}\gamma_5u|B^+\rangle.
\nonumber
\end{eqnarray}
The coefficients $L(\pi^0)$, $L(\pi^+)$, and $L(B^+)$ contain the coupling constants, colour factors, flavour symmetry factors, i.e. flavour
counting factors and factors resulting from the Fierz transformation of the four-quark operators from the effective weak Hamiltonian. The
coefficients $L(\pi^0)$ and $L(\pi^+)$ correspond to the quark decay diagram, whereas the $L(B^+)$ corresponds to the so-called
annihilation diagrams. These factors are different for
each decay mode, as indicated by the dependence on the final-state meson. To obtain the amplitudes for other decay modes, 
one has to replace the final-state particles with the particles relevent to that particular mode.
 
Finaly, we have to note that the sucessefull application of the factorization to the B decays was prooven, 
in the serious of works by the Benecke group \cite{benecke}. It was performed rigorousely, in the heavy quark limit, 
from the basic QCD principles.

We summarize all types of transitions in Table 4.

Next we present experimental results from the CLEO Collaboration \cite{cleononlept}
\begin{eqnarray}
&&BR(B^+\to K^+\pi^0)= \left(11.6 \;{+3.0 \atop -2.7} \;{+1.4 \atop -1.3}\right)\times 10^{-6},
\nonumber\\
&&BR(B^+\to K^0\pi^+)= \left(18.2 \;{+4.6 \atop -4.0} \;\pm 1.6\right)\times 10^{-6},
\nonumber\\
&&BR(B^0\to K^+\pi^-)= \left(17.2\;{+2.5 \atop -2.4} \;\pm 1.2\right)\times 10^{-6},
\nonumber\\
&&BR(B^0 \to K^0 \pi^0)= \left(14.6 \;{+5.9 \atop -5.1} \;{+2.4 \atop -3.3}\right)\times 10^{-6},
\\
&&BR(B^0\to \pi^+\pi^-)= \left(4.7 \;{+1.8 \atop -1.5} \;\pm 0.6\right)\times 10^{-6},
\nonumber\\
&&BR(B^{\pm}\to \pi^{\pm}\pi^0)= \left(5.4\; {+2.1 \atop -2.0}\; \pm 1.5\right)\times 10^{-6},
\\
&&BR(B^-\to \pi^- \rho^0)=\left(10.4 \;{+3.3 \atop -3.4} \;\pm 2.1\right)\times 10^{-6},
\nonumber\\
&&BR({\bar B}^0\to \pi^{\pm}\rho^{\mp})= \left(27.6 \;{+8.4 \atop -7.4} \;\pm 4.2\right)\times 10^{-6},
\nonumber\\
&&BR(B^-\to \pi^- \omega)= \left(11.3 \;{+3.3 \atop -2.9}\; \pm 1.4\right)\times 10^{-6},
\end{eqnarray}
which are in rough agreement with the very first theoretical attempts to 
predict the above measured rates \cite{guba,hou,dt3,tram}.

Recently from the $60\times 10^6$ $B{\bar B}$ pairs analysed from
$\Upsilon\,(4s)$, the BaBar Collaboration
published the following rates \cite{babar3}
\begin{eqnarray}
&&BR(B^0\to \pi^+\pi^-)= (5.4 \;\pm 0.7\; \pm 0.4)\times 10^{-6},
\nonumber\\
&&BR(B^0\to K^+\pi^-)= (17.8 \;\pm 1.1\; \pm 0.8)\times 10^{-6},
\nonumber\\
&&BR(B^0\to K^+K^-) < 1.1 \times 10^{-6} \;\; (90\% \;\rm CL),
\nonumber\\
&&BR(B^0\to \pi^0\pi^0) < 3.4 \times 10^{-6} \;\; (90\% \;\rm CL).
\end{eqnarray}

\renewcommand{\arraystretch}{1.8}
\begin{table}
\caption{The leading and subleading modes for $b \to s(d)\;{\bar q}q$ transitions.}
\begin{center}
\begin{tabular}{|c|c c c c c c|}
\hline 
 $ {B \to s{\bar q}q} $  & $\rm Leading  $ & $\rm Secondary $ & $\rm Sample $ &  $ B_d $ &  $\rm Sample $ & $ B_s $ \\
 $\rm modes $  & $\rm term $ & $\rm term $ & $ B_d\;\rm modes $ & $\rm angle $ & $ B_s \;\rm modes $ & $\rm angle $ \\
\hline\hline        
 $ { b\to s{\bar c}c} $  & $ V^*_{cs}V^{}_{cb}$ & $ V^*_{us}V^{}_{ub} $ & $ J/\psi K_S $ &  $ \beta $ &  $ \psi\eta $ & $ 0 $ \\
 $     $  & $\rm {tree \;+ \atop penguin(c-t)} $ & $\rm {only \atop penguin(u-t)} $ & $ \eta_c K $ & $  $ & $ D_s{\bar D}_s $ & $  $ \\
\hline         
 $ ^{\dagger}{ b\to s{\bar s}s} $  & $ V^*_{cs}V^{}_{cb}$ & $ V^*_{us}V^{}_{ub} $ & $ \phi K_S $ &  $ \beta $ &  $ \phi\eta' $ & $ 0 $ \\
 $     $  & $\rm {only \atop penguin(c-t)} $ & $\rm {only \atop penguin(u-t)} $ & $ \phi K^* $ & $  $ & $  $ & $  $ \\
\hline        
 $ { b\to s{\bar u}u} $  & $ V^*_{cs}V^{}_{cb}$ & $ V^*_{us}V^{}_{ub} $ & $ \pi^0 K_S $ &  $\rm competing $ &  $ \phi\pi^0 $ & $\rm competing $ \\
 $ { b\to s{\bar d}d} $  & $\rm {only \atop penguin(c-t)} $ & $\rm {tree \;+ \atop penguin(u-t)} $ & $ \rho K_S $ & $\rm terms $ & $ K_S{\bar K}_S $ & $ \rm terms $ \\
\hline\hline
 $ B \to d{\bar q}q   $  & $\rm Leading $ & $\rm Secondary $ & $\rm Sample $ &  $ B_d\; $ &  $\rm Sample $ & $ B_s $ \\
 $\rm modes $  & $\rm term $ & $\rm term $ & $ B_d \;\rm modes $ & $\rm angle $ & $ B_s \;\rm modes $ & $\rm angle $ \\
\hline\hline        
 $ { b\to d{\bar c}c} $  & $ V^*_{cd}V^{}_{cb}$ & $ V^*_{ud}V^{}_{tb} $ & $ D^+D^- $ &  $ {}^*\beta $ &  $ \psi K_S $ & $ {}^*\beta $ \\
 $     $  & $\rm {tree \;+ \atop penguin(c-u)} $ & $\rm {only \atop penguin(t-u)} $ & $  $ & $  $ & $ D_s{\bar D}_s $ & $  $
 \\
\hline         
 $ { b\to d{\bar s}s} $  & $ V^*_{td}V^{}_{tb}$ & $ V^*_{cd}V^{}_{cb} $ & $ \phi \pi $ &  $\rm competing $ &  $ \phi K_S $ & $\rm competing $ \\
 $     $  & $\rm {only \atop penguin(t-u)} $ & $\rm {only \atop penguin(c-u)} $ & $ K_S{\bar K}_S $ & $\rm terms $ & $  $ & $\rm terms $ \\
\hline        
 $ { b\to d{\bar u}u} $  & $ V^*_{ud}V^{}_{ub}$ & $ V^*_{td}V^{}_{tb} $ & $ \pi\pi;\;\pi\rho $ &  $ {}^*\alpha $ &  $\pi^0 K_S$ & $\rm competing $ \\
 $ { b\to d{\bar d}d} $  & $\rm {only \atop penguin(u-c)} $ & $\rm {tree \;+ \atop penguin(t-c)} $ & $ \pi \; a_1 $ & $  $ & $ \rho^0 K_S $ & $\rm terms $ \\
 \hline 
 $ {b \to c{\bar u}d} $  & $ V^*_{ud}V^{}_{cb} $ & $ 0 $ & $ D^0\pi^0,D^0\rho^0 $ &  $ \beta $ &  $ D^0 K_S $ & $ 0 $ \\
 $ $  & $  $ & $  $ & $\rm CP\; eigen\;st. $ & $  $ & $\rm CP \; eigen\;st. $ & $  $ \\
\hline\hline 
\end{tabular}\\
$~^*${\rm Leading terms only}.\\
$~^{\dagger}${\rm See analysis of CP asymmetry in Ref. \cite{deht}}.
\label{t:tab4}
\end{center}
\end{table}

They reconstructed a sample of B mesons $(B_{\rm rec})$ decaying to $\pi\pi$ and/or $\pi$K final states, 
and examine the remaining charged particles in each event to ``tag'' the flavour of the other B meson $(B_{\rm tag})$.
The decay rate distribution $f_+(f_-)$ in the case of $\pi^+\pi^-$ and $B_{\rm tag}=B^0({\bar B}^0)$ is given by
\begin{eqnarray}
f_{\pm}(\Delta t)=\frac{1}{4\tau} e^{-|\Delta t|/\tau}
\left[1\pm {\cal S}_{\pi\pi}\sin({\Delta m}_d \Delta t)\mp {\cal C}_{\pi\pi}\cos({\Delta m}_d \Delta t)\right],
\end{eqnarray}
where $\tau$ is the mean $B^0$ lifetime, ${\Delta m}_d$ is the eigenstate mass difference, and
$\Delta t = t_{\rm rec}-t_{\rm tag}$ is the time between the $(B_{\rm rec})$ and $(B_{\rm tag})$ decays. 
The asymmetry and $CP$-violating parameters
${\cal S}_{\pi\pi}$ and ${\cal C}_{\pi\pi}$ are defined as:
\begin{eqnarray}
{\cal A}_{K\pi}=\frac{N_{K^-\pi^+}-N_{K^+\pi^-}}{N_{K^-\pi^+} +N_{K^+\pi^-}},\; 
{\cal S}_{\pi\pi}=\frac{2\rm Im\lambda}{1+|\lambda|^2},\;
{\cal C}_{\pi\pi}=\frac{1-|\lambda|^2}{1+|\lambda|^2}.
\end{eqnarray}
The experimental results are
\begin{eqnarray}
{\cal A}_{K\pi}=-0.05\;\pm 0.06 \;\pm 0.01,
\\
{\cal S}_{\pi\pi}=-0.01\;\pm 0.37 \;\pm 0.07,
\nonumber\\
{\cal C}_{\pi\pi}=-0.02\;\pm 0.229 \;\pm 0.07.
\nonumber
\end{eqnarray}
For pure three diagram, through $b\to u\,W^-$ decay, we have 
\begin{eqnarray}
\lambda_{B\to \pi^+ \pi^-}=\frac{V^*_{tb}V_{td}^{}}{V_{tb}^{}V^*_{td}}\frac{V^*_{ud}V_{ub}^{}}{V_{ud}^{}V^*_{ub}}.
\end{eqnarray}
A small asymmetry ${\cal A}_{K\pi}$ disfavours many theoretical predictions and/or models
with large asymmetry.

\subsubsection{\it Exclusive and semi-inclusive B decays based on the $b\to s \eta_c$ transition}

The $b\to s \eta_c$ transition offers a unique opportunity to test our understanding of B meson decays.
The related process $b\to s \psi$ is known to give the ratio for semi-inclusive decays 
``$B\to \psi +$ anything'' to exclusive decays $B \to K\psi$ and $B\to K^*\psi$, in good agreement with data.
Here we show that by taking the ratio of processes involving $\eta_c$ to those involving $\psi$, one can 
remove the model dependence to a large extent, and have an independent and powerful way of determining 
$f_{\eta_c}$, the pseudoscalar decay constant of $\eta_c$, the $S_0$ state of charmonium.

The weak Hamiltonian corrected to NLO in QCD is given in subsection 1.2.1. The relevant QCD coefficients
we need are:
\begin{eqnarray}
&&c_1=1.150,\;\;\;\;\;c_2=-0.313,\;\;c_3=0.017,
\nonumber\\
&&c_4=-0.037,\;\;c_5=0.010,\;\;\; \;\;c_6=-0.046.
\end{eqnarray}
Now, we define the matrix elements
\begin{eqnarray}
\langle o|{\bar c}\gamma_{\mu}c|\psi(q)\rangle=i\epsilon_{\mu}(q)\;g_{\psi},\;
\langle o|{\bar c}\gamma_{\mu}\gamma_5 c|\eta_c(q)\rangle=iq_{\mu}\;f_{\eta_c},
\end{eqnarray}
where $g^2_{\psi}=(1.414\pm 0.083)$ GeV$^4$ from $\psi \to e^+e^-$ \cite{rpp}. The effective Hamiltonians in 
momentum space for the two decays are \cite{psi}:
\begin{eqnarray}
H_{b\to s\psi}^{eff}=\frac{G_F}{\sqrt 2}|V_{cs}^*V_{cb}^{}||C_{\psi}|g_{\psi}\epsilon_{\psi}^{\mu}(q)
{\bar s}^i(k)\gamma_{\mu}(1-\gamma_5)b^i(p),
\nonumber\\
H_{b\to s\eta_c}^{eff}=\frac{G_F}{\sqrt 2}|V_{cs}^*V_{cb}^{}||C_{\eta_c}|f_{\eta_c}q^{\mu}
{\bar s}^i(k)\gamma_{\mu}(1-\gamma_5)b^i(p),
\end{eqnarray}
where
\begin{eqnarray}
|C_{\psi}|=c_2+c_3 + c_5+\frac{1}{N_c} (c_1+c_4 + c_6),
\nonumber \\
|C_{\eta_c}|=c_2+c_3 - c_5+\frac{1}{N_c} (c_1+c_4 - c_6).
\end{eqnarray}
We shall treat $C_{\psi}$ and $C_{\eta_c}$ as phenomenological parameters, thus absorbing in their definition any higher-order 
correction or deviation from factorization that may arise. From Ref. \cite{etac} we use the stable ratio
$C_{\eta_c}/C_{\psi}=1.132\pm0.026$ and the $|C_{\psi}|=0.220\pm0.026$, which was determined following Ref. \cite{psi}.

The ratio of semi-inclusive $\psi$ production to $\eta_c$ production has been found to be
\begin{eqnarray}
\frac{\Gamma(B\to X_s\eta_c)}{\Gamma(B\to X_s\psi)}&\equiv&
\frac{\Gamma(b\to s\eta_c)}{\Gamma(b\to s\psi)}
\\
&=&\left|f_{\eta_c}\frac{C_{\eta_c}}{C_{\psi}}\right|^2\left(\frac{m_{\psi}}{g_{\psi}}\right)^2
\left(\frac{\lambda^a_{s\eta_c}}{\lambda^a_{s\psi}}\right)^{1/2}
\nonumber\\
&\times&\frac{(m^2_b-m^2_s)^2-m^2_{\eta_c}(m^2_b+m^2_s)}{m^2_b(m^2_b+m^2_{\psi})-m^2_s(2m^2_b-m^2_{\psi})+m^4_s-2m^4_{\psi}}
\nonumber\\
&\cong& 4.0(GeV^{-2})\left|f_{\eta_c}\frac{C_{\eta_c}}{C_{\psi}}\right|^2,
\nonumber
\end{eqnarray}
where $\lambda^a_{bc}=(1-m^2_b/m^2_a-m^2_c/m^2_a)^2-4m^2_bm^2_c/m^4_a$.
A measurement of this hadron-model-independent ratio offers a very accurate determination of $f_{\eta_c}$.

Next we consider the $B\to K\psi$ and $B\to K\eta_c$ exclusive modes. Using the general Lorentz decomposition 
of the vector current matrix element
\begin{eqnarray}
\langle K(k)|{\bar s}\gamma_{\mu} b|B(p)\rangle= (p+k)_{\mu} f^{(+)}_{KB}(q^2) + q_{\mu} f^{(-)}_{KB}(q^2),
\end{eqnarray}
we found the following ratio
\begin{eqnarray}
&&\frac{\Gamma(B\to K\eta_c)}{\Gamma(B\to K\psi)}= 
\left|f_{\eta_c}\frac{C_{\eta_c}}{C_{\psi}}\right|^2\left(\frac{m_{\psi}}{g_{\psi}}\right)^2
\frac{(\lambda^B_{K\eta_c})^{1/2}}{(\lambda^B_{K\psi})^{3/2}}
|f^{(+)}_{KB}(m_{\psi}^2)|^{-2}
\nonumber\\
&&\times \left|\left(1-\frac{m_K^2}{m_B^2}\right) f^{(+)}_{KB}(m_{\eta_c}^2) + 
 \frac{m_{\eta_c}^2}{m_B^2} f^{(-)}_{KB}(m_{\eta_c}^2)\right|^2.
\end{eqnarray}
Since $m_{\psi}\cong m_{\eta_c}$, we have set $f^{(+)}_{KB}(m_{\eta_c}^2)/f^{(+)}_{KB}(m_{\psi}^2)\cong 1$. The
second term in the above ratio is $\cong -0.06$, i.e. it is negligible with respect to the first term. An essentially
hadron-model-independent ratio is thus obtained:
\begin{eqnarray}
\frac{\Gamma(B\to K\eta_c)}{\Gamma(B\to K\psi)}\cong 14.2(GeV^{-2})
\left|f_{\eta_c}\frac{C_{\eta_c}}{C_{\psi}}\right|^2.
\end{eqnarray}
Finally, we calculate the exclusive rates for $B\to K^*\psi$ and $B\to K^*\eta_c$. Taking the general Lorentz decomposition
of the relevant (V--A) current from subsection 1.2.3, we obtain the hadron-model-dependent ratio
\begin{eqnarray}
&&\frac{\Gamma(B\to K^*\eta_c)}{\Gamma(B\to K^*\psi)}= 
\left|f_{\eta_c}\frac{C_{\eta_c}}{C_{\psi}}\right|^2\left(\frac{m_B + m_{K^*}}{g_{\psi}}\right)^2
\left(\frac{\lambda^B_{K^*\eta_c}}{\lambda^B_{K^*\psi}}\right)^{3/2}|A_0|^2
 \\
 &&\times \left[2|V|^2 + \left(\frac{3}{\lambda^B_{K^*\psi}}+\frac{m^4_B}{4m^2_{K^*}m^2_{\psi}}\right)
 \left(1-\frac{m^2_{K^*}}{m_B^2}\right)|A_1|^2 \right.
 \nonumber\\
 &&\left.+\left(\frac{m^4_B}{4m^2_{K^*}m^2_{\psi}}\right)\lambda^B_{K^*\eta_c}|A_2|^2 \right.
 \nonumber\\
&&\left.-\left(\frac{m^4_B}{2m^2_{K^*}m^2_{\psi}}\right)
\left(1-\frac{m^2_{\psi}}{m_B^2}-\frac{m^2_{K^*}}{m_B^2}\right)
\left(1-\frac{m^2_{K^*}}{m_B^2}\right)A_1 A_2 \right]^{-1}.
\nonumber
\end{eqnarray}
This ratio can be represented as 
\begin{eqnarray}
\frac{\Gamma(B\to K^*\eta_c)}{\Gamma(B\to K^*\psi)}= 
{\cal R} \left|f_{\eta_c}\frac{C_{\eta_c}}{C_{\psi}}\right|^2,
\end{eqnarray}
where the factor ${\cal R}$ depends on the hadronic model used. We consider a number of different models to
estimate that factor. Extensive discussion and various values of a factor 
${\cal R}$ are given in Ref. \cite{etac}. 

To estimate branching ratios for $B\to X_s\eta_c$, $B\to K\eta_c$ and $B\to K^*\eta_c$ decays, one has to know the pseudoscalar 
decay constant $f_{\eta_c}$. Theoretically, like the value of $g_{\psi}$, the quantity $f_{\eta_c}$ can be related to 
the wave function of the $S$-state of the charmonium at the origin:
\begin{eqnarray}
g^2_{\psi}=12m_{\psi}|\psi(0)|^2,\;\;f^2_{\eta_c}=48\frac{m^2_c}{m^2_{\eta_c}}|\psi(0)|^2.
\end{eqnarray}
Without QCD corrections the above expressions give $f_{\eta_c} \cong 350$ MeV. The QCD corrections are significant
but approximately cancel in the inclusive ratio.

A non-perturbative estimate of $f_{\eta_c}$ based on QCD sum rules \cite{rubi} could be more reliable. Following
Ref. \cite{rubi} we have found $f_{\eta_c}=(300\pm50)$ MeV. 

Using the central value of $f_{\eta_c}=300$ MeV, and taking the ratio $|C_{\eta_c}/C_{\psi}|=1.132$, we estimate the following 
branching ratios \cite{etac}:
\begin{eqnarray}
&&BR(B\to X_s\eta_c)=(4.61\pm1.15)\times10^{-3},
\nonumber\\
&&BR(B^-\to K^-\eta_c)=(1.80\pm0.29)\times10^{-3},
\nonumber\\
&&BR(B^0\to K^0\eta_c)=(1.23\pm0.41)\times10^{-3},
\nonumber\\
&&BR(B^-\to K^{*-}\eta_c)={\cal R}(0.21\pm0.07)\times10^{-3},
\nonumber\\
&&BR(B^0\to K^{*0}\eta_c)={\cal R}(0.20\pm0.04)\times10^{-3}.
\end{eqnarray}

In summary we have shown a very accurate technique of measuring $f_{\eta_c}$ from the measurement of relevant inclusive 
and exclusive branching ratios by predicting branching ratios of exclusive and inclusive ratios for 
the most important modes \cite{etac}. 
Note also that the measurement of $B\to K^*\eta_c$ probes the spin-0 part of the axial form factor and, again, 
provides a useful check of the model building.\\

The BaBar Collaboration presented, a few months ago the first measurements of the above branching ratios \cite{etacbab}. 
Their results,
\begin{eqnarray}
&&BR(B^+\to K^+\eta_c)=(1.50\pm0.19\pm0.15\pm0.46)\times10^{-3},
\nonumber\\
&&BR(B^0\to K^0\eta_c)=(1.06\pm0.29\pm0.11\pm0.33)\times10^{-3},
\end{eqnarray}
are almost perfectly placed within our predicted rates.

Taking the central BR values, 
from the charge and from the neutral decay mode measurements,
we obtain $f_{\eta_c} = 274$ MeV and $f_{\eta_c} = 279$ MeV, respectively. 

\subsection{\it Discussion and conclusions on the rare B meson decays}

As part of the discussion, I will first present interesting results on 
forward--backward asymmetry in ${\bar B}\to {\bar K}^* \mu^+\mu^-$
decay and the possibility that new physics arise through non-standard ${\bar b}sZ$ coupling \cite{buchalla1}:
\begin{eqnarray}
{\cal A}^{({\bar B})}_{FB}(s)&=&
\left(\frac{d\Gamma({\bar B}\to {\bar K}^* \mu^+\mu^-)}{ds}\right)^{-1}
\\
&\times &\int^1_{-1} d(\cos\theta) \frac{d^2\Gamma({\bar B}\to{\bar K}^* \mu^+\mu^-)}{ds d(\cos \theta)} {\rm sgn} (\cos\theta),
\nonumber
\end{eqnarray}
where, in the $\mu^+\mu^-$ c.m.s., the variable $s=m^2_{\mu^+\mu^-}/m^2_B$.

\begin{figure}
 \resizebox{0.8\textwidth}{!}{%
  \includegraphics{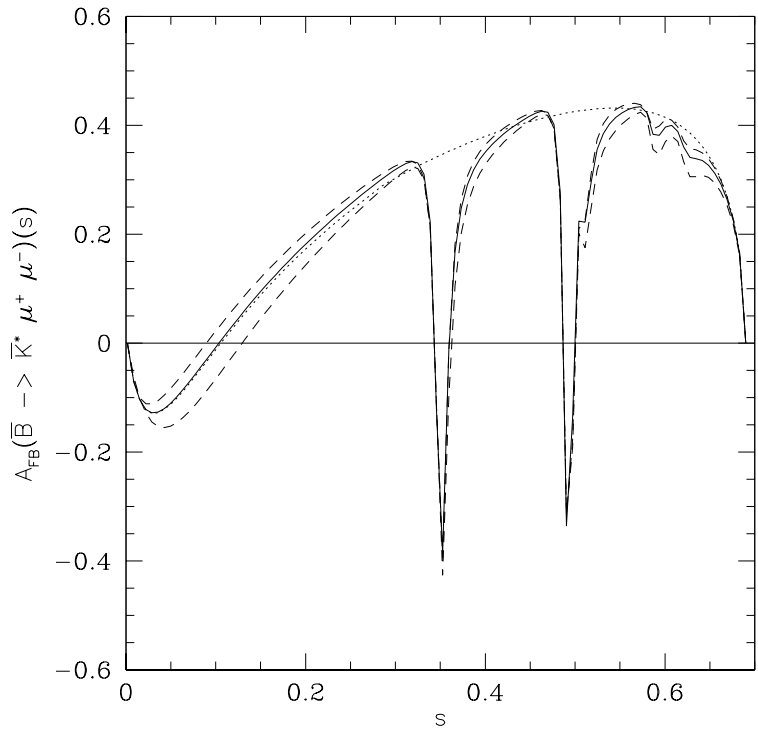}}
 \caption{The forward-backward asymmetry $A_{FB}({\bar B}\to {\bar K}^*\mu^+\mu^-)(s)$ 
 including nonperturbative effects from the resonant ${\bar c}c$ background \cite{buchalla1}.}
 \label{Zfig4}
\end{figure}

Since the lepton current has only (V--A) structure, then asymmetry ${\cal A}^{({\bar B})}_{FB}$ provides a direct measure
of the $A\diamondsuit V$ interference. The asymmetry after integration is proportional to
\begin{eqnarray}
{\cal A}^{({\bar B})}_{FB}(s)\sim \rm Re \left[C^*_{10}\left(sC_9^{eff}(s)+\alpha_+(s)\frac{m_b}{m_B}C_7\right)\right].
\end{eqnarray}
Fig. \ref{Zfig4}, from Ref. \cite{buchalla1}, shows very nicely the asymmetry as a function of the variable $s$.
From this the following conclusions could be drawn:\\
(1) in the case of CP conservation in the SM, we have ${\cal A}^{({\bar B})}_{FB}=-{\cal A}^{(B)}_{FB}$;\\
(2) because of hadronic uncertainties the ${\cal A}^{({\bar B})}_{FB}(s_0|_{SM}\cong 0.1)=0$ at the $10$\% level;\\
(3) in the SM ${\cal A}^{({\bar B})}_{FB}(s>s_0)>0$ and ${\rm sgn}C_{10}$ change in the presence of non-standard
${\bar b}sZ$ vertex, which is {\it the sign of new physics};\\
(4) the CP violation
\begin{eqnarray}
a_{\rm CP}({\bar B}\to {\bar K}^* \mu^+\mu^-)=
\frac{{\cal A}^{({\bar B})}_{FB}(s)+{\cal A}^{(B)}_{FB}(s)}{{\cal A}^{({\bar B})}_{FB}(s)-{\cal A}^{(B)}_{FB}(s)}
\end{eqnarray}
could rise up to $10$\% in the presence of new physics in the ${\bar b}sZ$ vertex;\\
(5) the resonant ${\bar c}c$ background was eliminated by taking the cut at $s<0.3$. The short-distance contributions
are then reduced by $\simeq 60$\% in agreement with Refs \cite{dt4,mis}.\\

A few important points have to be emphasized.

The so-called spectator-quark contributions \cite{don} 
and the first calculable non-perturbative,
essentially long-distance, correction \cite{vol}
to the inclusive rate are of the order of a few per cent. It has also been proved that
the fermionic (quarks and leptons) and photonic loop corrections to $b \to s \gamma$ reduce
$BR(b \to s \gamma)/BR(b \to ce\bar \nu)$ by $\sim 8\pm2$\% \cite{marc}.
Consequently, it is more appropriate to use $\alpha_{\rm em}=1/137$
for the real photon emission \cite{mis,marc}.\\

In general, we can conclude that, in theory, more effort is required in
calculating quark (inclusive) decays through higher loops.
A better understanding of bound states of heavy--light quarks (B meson etc.) 
and highly recoiled light quark bound states ($K^*$, $\rho$, \ldots)
is desirable.
This can be achieved by inventing new, more sophisticated perturbative
methods \cite{mel} and applying them
to the calculation of radiative B meson decays, which incorporate the full spectrum
of quark bound states ($K^*$, $\rho$, $K_1^*$, \ldots). In any case it looks like hadronic
form factors will be obtained in the future from basic-principle QCD calculations on the lattice.\\

In experiment, with a larger amount of data, we might expect a regular
but smaller and smaller increase of inclusive and exclusive branching ratios, and consequently 
stabilization of the hadronization rates: $R_{K^*}$, $R_{\rho}$, $R_{K^{**}}$, etc.; determinations
of $BR(B \to K_1^* \gamma)$ and some other, higher $K^{**}$ resonant modes; 
first measurements of $BR(b \to d \gamma)$, $BR(B \to \rho \gamma)$, and 
many other inclusive and exclusive rare B meson decay modes.

\newpage

\section{\it Forbidden decays}

On non-commutative space, the non-commutative Standard Model (NCSM) allows
new, usually SM-forbidden interactions: for exapmle, triple-gauge boson, 
fermion--fermion--2 gauge bosons interactions, 
photon coupling to left-handed and to sterile (right-handed) neutrinos, etc.
In these lectures we concentrate on 
decays, forbidden in the SM due to the Lorentz and gauge invariance.
They are $Z\rightarrow \gamma\gamma$ and $Z\rightarrow gg$ decays, 
from the gauge sector of the NCSM, 
the flavour-changing $K\to\pi\gamma$, 
$D\to(\pi,K)\gamma$, and $B\to(\pi,K,D)\gamma$ decays from the hadron sector,
and the ``transverse plasmon'' decay to neutrino antineutrino pairs, i.e. $\gamma_{\rm pl}\to \nu \bar\nu$.

For the gauge sector, it was necessary to construct the model, which we name it ``non-minimal NCSM'', which gives
the triple-gauge boson couplings. To consider plasmon decay we constructed the non-commutative Abelian action and 
estimate the rate $\Gamma(\gamma_{\rm pl}\to \nu \bar\nu)$.
For forbidden decays in the flavour-changing hadron sector, we constructed the effective, point-like, 
photon $\times$ current $\times$ current interaction
based on the minimal NCSM. The corrections due to the strong 
interactions are also taken into account.
The branching ratio for $K^+\to\pi^+\gamma$ decay estimated, in the static-quark approximation
and at a non-commutativity scale
of order 1/4 TeV, is predicted to be of the order of $10^{-16}$.

\subsection{\it Introduction to the non-commutative gauge theories}

The idea that coordinates may not commute 
can be traced back to Heisenberg.
A simple way to introduce a non-commutative structure into
spacetime is to promote the usual spacetime 
coordinates $x$ to non-commutative (NC) coordinates 
$\hat x$ with~\cite{sny}--\cite{DH}
\begin{equation}
\left[{\hat x}^{\mu},{\hat x}^{\nu} \right]=i\theta^{\mu\nu}, \quad
\left[\theta^{\mu\nu},{\hat x}^{\rho} \right]=0,                \label{CR}
\end{equation}
where $\theta^{\mu\nu}$ is a constant, real, antisymmetric matrix.
The non-commutativity scale $\Lambda_{NC}$ is fixed
by choosing dimensionless matrix elements 
$c^{\mu\nu}=\Lambda_{NC}^2\,
\theta^{\mu\nu}$ of order 1.
The original motivation to study such a scenario
was the hope that the introduction
of a fundamental scale
could deal with the infinities of
quantum field theory in a natural way.

Apart from many technical merits, the possibility of 
a non-commutative structure 
of space-time is of interest in its own right,
and its experimental discovery would be a result of fundamental importance. 

Note that the commutation relation (\ref{CR}) 
enters in string theory through
the Moyal--Weyl star product
\begin{equation}
f \star g = \sum_{n=0}^\infty \frac{\theta^{\mu_1 \nu_1} 
\cdots \theta^{\mu_n \nu_n}}{(-2i)^nn!}  
\partial_{\mu_1}\ldots\partial_{\mu_n} f
\cdot\partial_{\nu_1}\ldots\partial_{\nu_n} g,
\end{equation} 
which for coordinates is: $x^\mu \star x^\nu - x^\nu \star x^\mu = i \theta^{\mu\nu}$.

Experimental signatures of non-commutativity have been discussed from the point of 
view of collider physics~\cite{AY}--\cite{MPR} as 
well as low-energy non-accelerator experiments~\cite{MPR}--\cite{CCL}. 
Two widely disparate sets of bounds on $\Lambda_{NC}$
can be found in the literature: bounds of order $10^{11}$ $GeV$~\cite{ABDG} 
or higher~\cite{MPR}, and 
bounds of a few TeV from colliders~\cite{AY}--\cite{HK}.
All these limits rest on one or more of the following assumptions, which 
may have to be modified:\\ 
(a) $\theta$ is constant across distances that are large 
with respect to the NC scale; \\
(b) unrealistic gauge groups; \\
(c) non-commutativity down to low-energy scales.

Non-commutative gauge field theory (NCGFT) as it appears 
in string theory is, strictly speaking, limited to
the case of $U(N)$ gauge groups, where the Seiberg--Witten (SW) map~\cite{SW} 
plays an essential role since it does
express non-commutative gauge fields in terms of fields
with ordinary ``commutative'' gauge transformation properties. 
A method of constructing models on non-commutative space-time
with more realistic gauge groups and particle content
has been developed in a series of papers by the 
Munich group~\cite{WESS} and~\cite{Zumino},
culminating in the construction of the NCSM \cite{cal}. 

This construction for a given NC space rests on few basic ideas,
which it was necessary to incorporate \cite{ws}: \\
(1) non-commutative coordinates,\\
(2) the Moyal--Weyl star product,\\
(3) enveloping algebra-valued gauge transformation has to be used,\\
(4) Seiberg--Witten map as a most important new idea.\\
(5) Concepts of covariant coordinates, locality, gauge equivalence, and
consistency conditions had to be maintained.

The problems that are solved in this approach include, in addition to the
introduction of general gauge groups, 
the charge quantization problem of NC Abelian gauge theories
and the construction of covariant Yukawa couplings.
 
There are two essential points in which NC gauge theories differ 
from standard gauge theories. 
The first point is the breakdown of Lorentz invariance with
respect to a fixed non-zero $\theta^{\mu\nu}$ background 
(which obviously fixes preferred directions) the second
is the appearance of new interactions 
(triple-photon coupling, for example) and the modification of standard ones. 
Both properties have a common origin and appear in a number of phenomena. 

\subsection{\it Non-commutative standard model decays}
\subsubsection{\it Gauge sector: $Z\to \gamma\gamma,\;\,gg$ decays}

Strictly SM forbidden decays coming from the gauge sector of the NCSM
could be probed in high energy collider experiments.
This sector is particularly 
interesting from the theoretical point of view. 
Our main results are summarized in (\ref{eqn0}) to (\ref{eqn3}).\\

The general form of the gauge-invariant action 
for gauge fields is \cite{cal}
\begin{equation}
S_{gauge} =-\frac{1}{2}\int d^4x {\bf Tr}\frac{1}{{\bf G}^2}
{\widehat F}_{\mu\nu} \star {\widehat F}^{\mu\nu}.
\end{equation}
Here ${\bf Tr}$ is a trace 
and ${\bf G}$ is an operator that encodes the
coupling constants of the theory. Both will be discussed in detail below.
The NC field strength is 
\begin{equation}
{\widehat F}_{\mu\nu} = \partial_{\mu} {\widehat V}_{\nu} - \partial_{\nu} {\widehat V}_{\mu}
- i[{\widehat V}_{\mu}\stackrel{\star}{,}{\widehat V}_{\nu}]
\end{equation}
and ${\widehat V}_{\mu}$ is the NC analogue of the gauge vector potential. 
The Seiberg--Witten maps are used to express the non-commutative fields and 
parameters as functions of 
ordinary fields and parameters and their derivatives. 
This automatically ensures a restriction to the correct degrees of freedom.
For the NC vector potential, the SW map yields
\begin{equation}
{\widehat V}_{\xi}=V_{\xi}+\frac{1}{4}{\theta}^{\mu\nu}
\{V_{\nu},(\partial_{\mu}V_{\xi}+F_{\mu\xi})\}+{\cal O}\left(\theta^2 \right),
\end{equation}
where $F_{\mu\nu}\equiv \partial_{\mu}V_{\nu} - \partial_{\nu}V_{\mu} - i[V_{\mu},V_{\nu}]$
is the ordinary field strength and
$V_{\mu}$ is the whole gauge potential for the 
gauge group $G_{SM}\equiv SU(3)_C \times SU(2)_L \times U(1)_Y$:
\begin{equation}
V_{\mu}=g'{\cal A}_{\mu}(x)Y + g\sum^3_{a=1}B_{\mu,a}(x)T^a_L + g_s\sum^8_{b=1}G_{\mu,b}(x)T^b_S.
\end{equation}
It is important to realize that the choice of the representation
in the definition of the trace ${\bf Tr}$
has a strong influence on the theory in the non-commutative case.
The reason for this is that, owing to the Seiberg--Witten map, terms
of higher than quadratic order in the Lie algebra generators will 
appear in the trace.
 The adjoint representation as, a
natural choice for the non-Abelian gauge fields, shows no
triple-photon vertices~\cite{cal,ASCH}.\\

The action that we present here
should be understood as an effective theory. 
According to \cite{cal}, we choose a trace over all particles 
with different quantum numbers in the model that have
covariant derivatives acting on them.
In the SM, these are, for each generation, 
five fermion multiplets and one Higgs multiplet. 
The operator ${\bf G}$, which determines the coupling constants of the theory,
must commute with all generators 
$(Y,T^a_L,T^b_S)$ of the gauge group,
so that it does not spoil the trace property of ${\bf Tr}$, i.e. the ${\bf G}$ 
takes on constant values $g_1,\ldots,g_6$
on the six multiplets (Table 1 in Ref.~\cite{cal}).\\

The action up to linear order in $\theta$ 
allows new triple gauge boson interactions that are forbidden in the SM
and has the following form \cite{behr}:
\begin{eqnarray}
\lefteqn{S_{gauge}=-\frac{1}{4}\int \hspace{-1mm}d^4x\, f_{\mu \nu} f^{\mu \nu}}
 \label{action2} \\
& &\hspace{-5mm}{}
-\frac{1}{2}\int \hspace{-1mm}d^4x\, {\rm Tr}\left( F_{\mu \nu} F^{\mu \nu}\right)
-\frac{1}{2}\int\hspace{-1mm} d^4x\, {\rm Tr}\left( G_{\mu \nu} G^{\mu \nu}\right)
\nonumber \\
& &\hspace{-5mm}{}
+g_s \,\theta^{\rho\tau}\hspace{-2mm}
\int\hspace{-1mm} d^4x\, {\rm Tr}
\left(\frac{1}{4} G_{\rho \tau} G_{\mu \nu} - G_{\mu \rho} G_{\nu \tau}\right)G^{\mu \nu}\nonumber \\
& &\hspace{-5mm}{}+{g'}^3\kappa_1{\theta^{\rho\tau}}\hspace{-2mm}\int \hspace{-1mm}d^4x\,
\left(\frac{1}{4}f_{\rho\tau}f_{\mu\nu}-f_{\mu\rho}f_{\nu\tau}\right)f^{\mu\nu}
 \nonumber \\
& &\hspace{-5mm}{}+g'g^2\kappa_2 \, \theta^{\rho\tau}\hspace{-2mm}\int
\hspace{-1mm} d^4x \sum_{a=1}^{3}
\left[(\frac{1}{4}f_{\rho\tau}F^a_{\mu\nu}-
f_{\mu\rho}F^a_{\nu\tau})F^{\mu\nu,a}\!+c.p.\right]
 \nonumber \\
& &\hspace{-5mm}{}+g'g^2_s\kappa_3\, \theta^{\rho\tau}\hspace{-2mm}\int
\hspace{-1mm} d^4x \sum_{b=1}^{8}
\left[(\frac{1}{4}f_{\rho\tau}G^b_{\mu\nu}-
f_{\mu\rho}G^b_{\nu\tau})G^{\mu\nu,b}\!+c.p.\right], \nonumber 
\end{eqnarray}
where $c.p.$ means cyclic permutations in $f$.
Here $f_{\mu\nu}$, $F^a_{\mu\nu}$, and $G^b_{\mu\nu}$ are the physical field strengths corresponding 
to the groups $U(1)_Y$, $SU(2)_L$, and $SU(3)_C$, respectively. 
The constants $\kappa_1$, $\kappa_2$, and $\kappa_3$ are functions of $1/g_i^2\; (i=1,...,6)$ 
and have the following form:
\begin{eqnarray}
\kappa_1 &=& -\frac{1}{g^2_1}-\frac{1}{4g^2_2}+\frac{8}{9g^2_3}-\frac{1}{9g^2_4}+\frac{1}{36g^2_5}
+\frac{1}{4g^2_6},
\nonumber \\
\kappa_2 &=& -\frac{1}{4g^2_2}+\frac{1}{4g^2_5}+\frac{1}{4g^2_6},
\nonumber \\
\kappa_3 &=& +\frac{1}{3g^2_3}-\frac{1}{6g^2_4}+\frac{1}{6g^2_5}.
\end{eqnarray}
In order to match the SM action at zeroth order in $\theta$, three consistency conditions
have been imposed in (\ref{action2}):
\begin{eqnarray}
\frac{1}{{g'}^2} &=& \frac{2}{g^2_1}+\frac{1}{g^2_2}+\frac{8}{3g^2_3}+\frac{2}{3g^2_4}+\frac{1}{3g^2_5}
+\frac{1}{g^2_6},
\nonumber \\
\frac{1}{g^2}&=& \frac{1}{g^2_2}+\frac{3}{g^2_5}+\frac{1}{g^2_6},\nonumber \\
\frac{1}{g_s^2}&=& \frac{1}{g^2_3}+\frac{1}{g^2_4}+\frac{2}{g^2_5}.
\end{eqnarray}
These three conditions, together with the requirement that 
$1/g_i^2 > 0$, define a three-dimensional simplex in
the six-dimensional moduli space spanned by $1/g_1^2,...,1/g_6^2$. 
Since the last three couplings in (\ref{action2}) are 
not uniquely fixed by the NCSM, they need to be determined through the 
various types of physical processes, such as decays and collisions, unpolarized-polarized, etc.\\

\begin{figure}
\resizebox{0.55\textwidth}{!}{\includegraphics{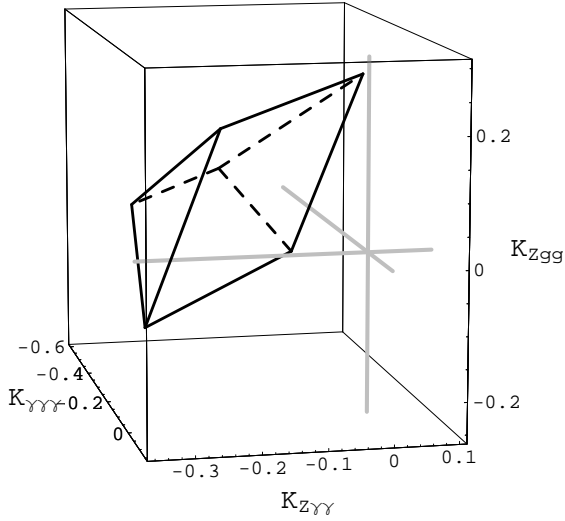}}
\caption{ The three-dimensional simplex that bounds possible
values for the coupling constants K$_{\gamma\gamma\gamma}$,
K$_{Z\gamma\gamma}$ and K$_{Zgg}$ at the  $M_Z$ scale. The
vertices of the simplex are:
 ($-0.184$, $-0.333$, $0.054$), ($-0.027$, $-0.340$, $-0.108$),
 ($0.129$, $-0.254$, $0.217$),  ($-0.576$, $0.010$, $-0.108$),
 ($-0.497$, $-0.133$, $0.054$) and ($-0.419$, $0.095$, $0.217$).
}
\label{fig1}
\end{figure}

From the action (\ref{action2}) we extract the 
neutral triple-gauge boson terms which are not present in the SM Lagrangian. 
In terms of physical fields ($A,Z,G$) they are \cite{behr}
\begin{eqnarray}
{\cal L}_{\gamma\gamma\gamma}&=&\frac{e}{4} \sin2{\theta_W}\;{\rm K}_{\gamma\gamma\gamma}
{\theta^{\rho\tau}}A^{\mu\nu}\left(A_{\mu\nu}A_{\rho\tau}-4A_{\mu\rho}A_{\nu\tau}\right),\nonumber\\
{\rm K}_{\gamma\gamma\gamma}&=&\frac{1}{2}\; gg'(\kappa_1 + 3 \kappa_2);  \label{L1}\\
& & \nonumber \\
{\cal L}_{Z\gamma\gamma}&=&\frac{e}{4} \sin2{\theta_W}\,{\rm K}_{Z\gamma \gamma}\,
{\theta^{\rho\tau}}
\left[2Z^{\mu\nu}\left(2A_{\mu\rho}A_{\nu\tau}-A_{\mu\nu}A_{\rho\tau}\right)\right.\nonumber\\
& & +\left. 8 Z_{\mu\rho}A^{\mu\nu}A_{\nu\tau} - Z_{\rho\tau}A_{\mu\nu}A^{\mu\nu}\right], \nonumber \\
{\rm K}_{Z\gamma\gamma}&=&\frac{1}{2}\; \left[{g'}^2\kappa_1 + \left({g'}^2-2g^2\right)\kappa_2\right]; \label{L2}\\
& &\nonumber \\
{\cal L}_{ZZ\gamma}&=&{\cal L}_{Z\gamma\gamma}(A\leftrightarrow Z),\nonumber \\
{\rm K}_{ZZ\gamma}&=&\frac{-1}{2gg'}\; \left[{g'}^4\kappa_1 + g^2\left(g^2-2{g'}^2\right)\kappa_2\right]; \label{L3}\\
& &\nonumber \\
{\cal L}_{ZZZ}&=&{\cal L}_{\gamma\gamma\gamma}(A\to Z),\nonumber\\
{\rm K}_{ZZZ}&=&\frac{-1}{2g^2}\; \left[{g'}^4\kappa_1 + 3g^4\kappa_2\right]; \label{L4}\\
& &\nonumber \\
{\cal L}_{Zgg}&=&{\cal L}_{Z\gamma\gamma}(A\to G^b), \nonumber \\
{\rm K}_{Zgg}&=&\frac{g^2_s}{2} \left[1+(\frac{{g'}}{g})^2\right]\kappa_3; \label{L5}\\
& &\nonumber \\
{\cal L}_{\gamma gg}&=&{\cal L}_{Zgg}(Z\rightarrow A), \nonumber \\
{\rm K}_{\gamma gg}&=&\frac{-g^2_s}{2}\;
\left[\frac{g}{g'}+\frac{g'}{g}\right]\kappa_3, \label{L6}
\end{eqnarray} 
where $A_{\mu\nu} \equiv \partial_{\mu}A_{\nu} -
\partial_{\nu}A_{\mu}$, etc.
Fig. \ref{fig1} shows the three-dimensional simplex that bounds 
allowed values for the dimensionless coupling constants
${\rm K}_{\gamma\gamma\gamma}$, ${\rm K}_{Z\gamma\gamma}$ 
and ${\rm K}_{Zgg}$. For any choosen point within simplex 
in Fig. \ref{fig1}, the remaining three coupling constants (\ref{L3}, \ref{L4}, \ref{L6}), i.e.
${\rm K}_{Z Z \gamma}$, ${\rm K}_{Z Z Z}$ 
and ${\rm K}_{\gamma g g}$ respectively, are uniquely fixed by the NCSM.
This is true for any combination of three coupling constants from Eqs. (\ref{L1}) to (\ref{L6}).

Experimental evidence for non-commutativity coming from the gauge sector
that should be searched for 
in processes involve the above couplings. 
The simplest and most natural choice are the 
$Z\rightarrow \gamma\gamma, \;gg$ decays, allowed for real (on-shell) particles.
All other simple processes, such as
$\gamma \rightarrow \gamma \gamma, \;gg$, and $Z\rightarrow Z\gamma, \;ZZ$,
are on-shell-forbidden by kinematics. 
The $Z\rightarrow \gamma\gamma, \: gg$ decays are strictly
forbidden in the SM by Lorentz and gauge invariance; 
both could therefore serve as a clear signal for the existence of 
space-time non-commutativity.\\

There is huge interest among the experimentalists to find the anomalous triple-gauge boson
couplings \cite{LEPEWWG}, since such observation would certainly contribute to the
discovery of physics beyond the SM.
The experimental upper bound, obtained from the $e^+e^-\rightarrow \gamma\gamma$ annihilation, 
for $Z \rightarrow \gamma\gamma$, is:
\begin{eqnarray}
\Gamma(Z \rightarrow \gamma\gamma) \;< \;\left(\begin{array}{c}5.2\\5.5\\14.0
\end{array}\right)\times 10^{-5} \;{\rm GeV},
\left(\begin{array}{c}{\rm from \;L3\; \cite{L3}}\\{\rm from\; DELPHI\; \cite{DELPHI}}\\{\rm from \;OPAL \;\cite{OPAL}}
\end{array}\right).
\end{eqnarray}
\noindent
Note that the $Z\rightarrow \gamma\gamma$ process has a tiny SM 
background from the rare $Z\rightarrow \pi^0\gamma,\;\eta\gamma$ decays. 
At high energies, the two photons from the $\pi^0$ or $\eta$ decay
are too close to be separated and they are seen in the electromagnetic calorimeter as 
a single high-energy photon \cite{EXP}. The SM
branching ratios for these rare decays are of order $10^{-11}$
to $10^{-10}$ \cite{ALT}. This is much smaller than the experimental upper bounds
which are of order $10^{-5}$ for the all three branching ratios 
($Z\rightarrow \gamma\gamma,\; \pi^0\gamma,\; \eta\gamma$) \cite{rpp}. \\

The $Z\rightarrow gg$ decay mode should be observed in $Z\rightarrow 2\;{\rm jets}$ processes.
However, it could be smothered by the strong 
$Z\rightarrow q{\bar q}$ background, i.e. by hadronization, which also
contains NC contributions. Since
the hadronic width of the $Z$ is in good agreement with the QCD-corrected SM, 
the $Z\rightarrow gg$ can be at most a few per cent.
Taking into account the discrepancy between the experimentally 
observed hadronic width for the $Z$-boson 
and the theoretical estimate based on the radiatively corrected SM, 
we estimate the upper bound for any new hadronic
mode, such as $\Gamma_{Z \rightarrow gg}$ to be $\sim 10^{-3}$ GeV \cite{rpp}.\\

We now derive the partial widths for the $Z(p) \rightarrow \gamma (k)\,\gamma (k')$ decay.
From the Lagrangian ${\cal L}_{Z\gamma \gamma}$, it is easy to write 
the gauge-invariant amplitude ${\cal M}_{Z\rightarrow \gamma\gamma}$ in momentum space, which gives: 
\begin{eqnarray}
\sum_{spins}\,|{\cal M}_{Z\rightarrow \gamma \gamma}|^2 
= -{\theta}^2 + \frac{8}{M^2_Z}(p{\theta}^2 p)
- \frac{16}{M^4_Z}(k{\theta}k')^2 \, .
\label{eqn0}
\end{eqnarray}
From the above equation and in the $Z$-boson rest frame, the partial width
of the $Z \rightarrow \gamma\gamma$ decay is \cite{behr}:
\begin{equation}
\Gamma_{Z\rightarrow \gamma\gamma} 
= \frac{\alpha}{12} \frac{M^5_Z}{\Lambda^4_{NC}} \sin^2 2\theta_W {\rm K}^2_{Z\gamma \gamma} 
\left[\frac{7}{3}({\vec {E_{\theta}}})^2+({\vec {B_{\theta}}})^2\right],
\label{eqn1}
\end{equation}
where ${\vec {E_{\theta}}}=(c^{01},c^{02},c^{03})$ 
and ${\vec {B_{\theta}}}=(c^{23},c^{13},c^{12})$, are
responsible for time--space and space--space non-commutativity, respectively. 
This result differs essentially from that given in \cite{MPR},
where the $\Gamma_{Z\rightarrow \gamma\gamma}$
partial width depends only on time--space non-commutativity.

For the $Z$-boson at rest and polarized in the 
direction of the $3$-axis, we find that the \emph{polarized} partial width is \cite{behr}
\begin{eqnarray}
& &\Gamma_{Z^3 \rightarrow \gamma\gamma }\;=\;
\frac{\alpha}{4} \;\frac{M^5_Z}{\Lambda^4_{NC}} \;\sin^2 2\theta_W\;{\rm K}^2_{Z\gamma \gamma} 
\nonumber \\
& &\times \left[\frac{2}{5}
\left((c^{01})^2+(c^{02})^2\right)
+\frac{23}{15}(c^{03})^2+(c^{12})^2\right]. \label{eqn2}
\end{eqnarray}
In the absence of time--space non-commutativity 
a sophisticated, sensibly arranged  polarization
experiment could in principle determine the vector of ${\vec {E_{\theta}}}$. 
A NC structure of space-time may depend on the matter that is present. 
In our case it is conceivable that the direction of ${\vec {E_{\theta}}},\;{\vec {B_{\theta}}}$
may be influenced by the polarization of the $Z$ particle.
In this case, our result for the \emph{polarized} partial width is particularly relevant.

Since the Lagrangians ${\cal L}_{Z\gamma\gamma}$ and ${\cal L}_{Zgg}$ have the same Lorentz structure, 
we find
\begin{eqnarray}
\frac{\Gamma_{Z\rightarrow gg}}{\Gamma_{Z\rightarrow \gamma\gamma}}\;=\;
\frac{\Gamma_{Z^3\rightarrow gg}}{\Gamma_{Z^3\rightarrow \gamma\gamma}}\;=\;
8\frac{{\rm K}^2_{Zgg}}{{\rm K}^2_{Z\gamma \gamma}}. \label{eqn3}
\end{eqnarray}
The factor of 8 in the above ratios is due to colour.

In order to estimate the NC parameter from upper bounds 
$\Gamma^{exp}_{Z \rightarrow \gamma\gamma} < 1.3 \times 10^{-4}$ GeV and
$\Gamma^{exp}_{Z \rightarrow gg} < 1 \times 10^{-3}$ GeV \cite{rpp}
it is necessary to determine the range of couplings ${\rm K}_{Z\gamma\gamma}$ and ${\rm K}_{Zgg}$.
\begin{figure}
 \resizebox{0.6\textwidth}{!}{%
  \includegraphics{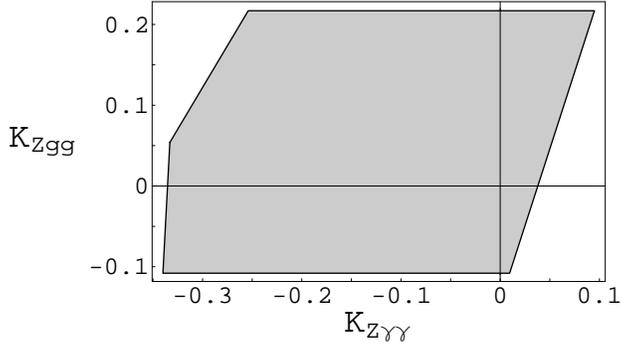}}
 \caption{The allowed region for ${\rm K}_{Z\gamma\gamma}$ and ${\rm K}_{Zgg}$ 
at the $M_Z$ scale, projected from the simplex given in Fig.5. The vertices of the polygon are
$(-0.254,\, 0.217)$, 
$(-0.333,\, 0.054)$, 
$(-0.340,\, -0.108)$, 
$(0.010,\, -0.108)$ and 
$(0.095, \,0.217)$.}
 \label{fig2a}
\end{figure}
\begin{figure}
 \resizebox{0.6\textwidth}{!}{%
  \includegraphics{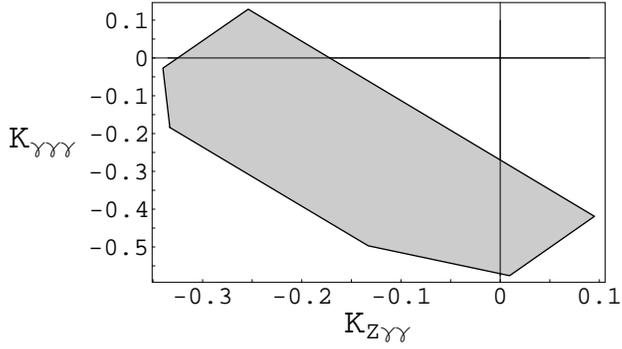}}
 \caption{The allowed region for ${\rm K}_{Z\gamma\gamma}$ 
 and ${\rm K}_{\gamma\gamma\gamma}$  at the $M_Z$ scale, projected from
 the simplex given in Fig.5. The vertices of the polygon are
 $(-0.333,\, -0.184)$, $(-0.340,\, -0.027)$, $(-0.254,\, 0.129)$, $(0.095,\, -0.419)$, 
 $(0.0095, \,-0.576)$, and $(-0.133,\, -0.497)$.}
 \label{fig2b}
\end{figure}
\begin{figure}
 \resizebox{0.6\textwidth}{!}{%
  \includegraphics{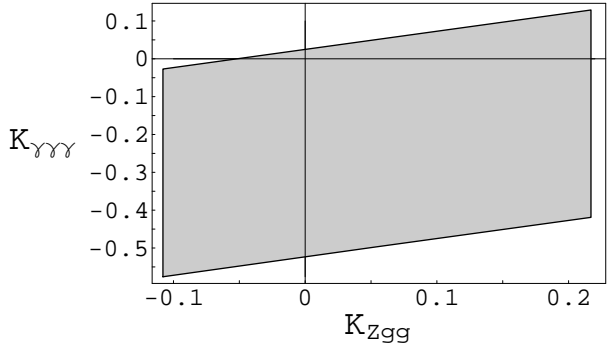}}
 \caption{The allowed region for ${\rm K}_{\gamma\gamma\gamma}$
 and ${\rm K}_{Zgg}$ at the $M_Z$ scale, projected from the simplex
 given in Fig.5. The vertices of the polygon are $(-0.108,\, -0.576)$, $(-0.108,\, -0.027)$, 
 $(0.217,\, 0.129)$, $(0.217,\, -0.419)$, and $(0.054,\, -0.497)$.}
 \label{fig2c}
\end{figure}
The allowed region for the coupling constants ${\rm K}_{Z\gamma\gamma}$ and ${\rm K}_{Zgg}$ 
is given in Fig. \ref{fig2a}.
Since ${\rm K}_{Z\gamma\gamma}$ and ${\rm K}_{Zgg}$ could be zero
simultaneously, it is not possible to extract an upper bound on $\theta$ 
from the above experimental upper bounds alone.\\

To succeed in estimating $\theta$, we 
should consider an extra interaction from the NCSM gauge sector, in
particular triple-photon vertices. 
From the simplex we find that the triplet of coupling constants 
${\rm K}_{\gamma\gamma\gamma}$, ${\rm K}_{Z\gamma\gamma}$ and ${\rm K}_{Zgg}$,
as well as the pair of couplings ${\rm K}_{\gamma\gamma\gamma}$ and ${\rm K}_{Z\gamma\gamma}$,
{\it cannot vanish simultaneously} (see e.g. Fig. \ref{fig2b}) 
and that it is possible to estimate $\theta$ 
from the NCSM gauge sector through a combination of various types of
processes containing the $\gamma\gamma\gamma$ and $Z\gamma\gamma$ vertices.
These are processes of the type $2 \rightarrow 2$, such as 
$e^+e^-\rightarrow \gamma\gamma$, 
$e\gamma \rightarrow e\gamma$, and $\gamma\gamma \rightarrow e^+e^-$ 
in leading order. Such inclusion of other triple-gauge boson interactions sufficientlly reduce
available parameter space.
The analysis has to be carried out in the same way as in Ref. \cite{HPR}.
Theoretically consistent modifications of relevant vertices are, however, necessary. 
The allowed region for pairs of couplings ${\rm K}_{\gamma\gamma\gamma}$ and
${\rm K}_{Zgg}$ is presented in Fig. \ref{fig2c}. 

\subsubsection{\it Hadron sector -- flavour changing decays: $K\to \pi\gamma$, ...}

From the action (55) in Ref. \cite{cal}, for quarks that couples to an non-Abelian gauge boson in a 
non-commutative background, we obtain the explicit formulas for the electroweak 
charged currents in the leading order of the expansion in $\theta$:
\begin{equation}
{\cal L}_{CC} =\left(\begin{array}{ccc}
{\bar u}\;\;{\bar c}\;\;{\bar t}
\end{array}\right)_L  J_+ V_{CKM}  
\left(\begin{array}{c}
d \\ s \\ b
\end{array}\right)_L 
+ \left(\begin{array}{ccc}
{\bar d}\;\;{\bar s}\;\;{\bar b}
\end{array}\right)_L  J_- V^*_{CKM} 
\left(\begin{array}{c}
u \\ c \\ t
\end{array}\right)_L,
\label{1}
\end{equation}
were $J_{\pm}$ are given in eqs. (72) and (73) of Ref. \cite{cal}. 
Note that for left-handed quarks the hypercharge $Y=1/6$.

Isolating terms linear in {\it W} and {\it A} fields, we have found the following charged current:
\begin{eqnarray}
&&{\bar\psi}J_+\psi' = \frac{g}{\sqrt 2}{\bar\psi}\gamma^{\mu}W^+_{\mu}\psi' - \frac{g^2}{\sqrt 2}\sin\theta_w
{\bar\psi}\left( \frac{1}{2}\theta^{\mu\nu}\gamma^{\alpha}+\theta^{\nu\alpha}\gamma^{\mu}\right)
\nonumber \\
&& \times \left[\frac{1}{3}
\left( A_{\mu\nu}W^+_{\alpha}-\frac{1}{2}W^+_{\mu\nu}A_{\alpha}\right)
+\left(A_{\mu}W^+_{\nu}-A_{\nu}W^+_{\mu} \right)\partial_{\alpha}\right]\psi',
\label{2c}
\end{eqnarray}
with $A_{\mu\nu}\equiv \partial_{\mu}A_{\nu}- \partial_{\nu}A_{\mu}$, etc.\\

To simplify the calculation of the $K\to \pi \gamma$ decay rate, I use the static-quark approximation (sqa)
in the following way.

First we modify the charged current 
by applying the integration by parts on the ${\bar\psi}W^+_{\mu\nu}A_{\alpha}\psi$ term of the above equation.
We then use the static-quark approximation in the above equation by neglecting all derivatives acting on quark fields, 
i.e. by putting $\partial_{\alpha}\psi = \partial_{\alpha}{\bar\psi}=0$, and obtain the following expression:
\begin{eqnarray}
&&{\bar\psi}J_+\psi' = \frac{g}{3\sqrt 2}{\bar\psi}\gamma^{\mu}W^+_{\mu}\psi' - \frac{g^2}{\sqrt 2}\sin\theta_w
{\bar\psi}\left(\frac{1}{2}\theta^{\mu\nu}\gamma^{\alpha}+\theta^{\nu\alpha}\gamma^{\mu}\right)
\nonumber \\
&& \times \left[ A_{\mu\nu}W^+_{\alpha}-\frac{1}{2}
\left((\partial_{\nu}A_{\alpha})W^+_{\mu}-(\partial_{\mu}A_{\alpha})W^+_{\nu}\right)\right]\psi'.
\end{eqnarray}
\begin{figure}
 \resizebox{0.9\textwidth}{!}{%
  \includegraphics{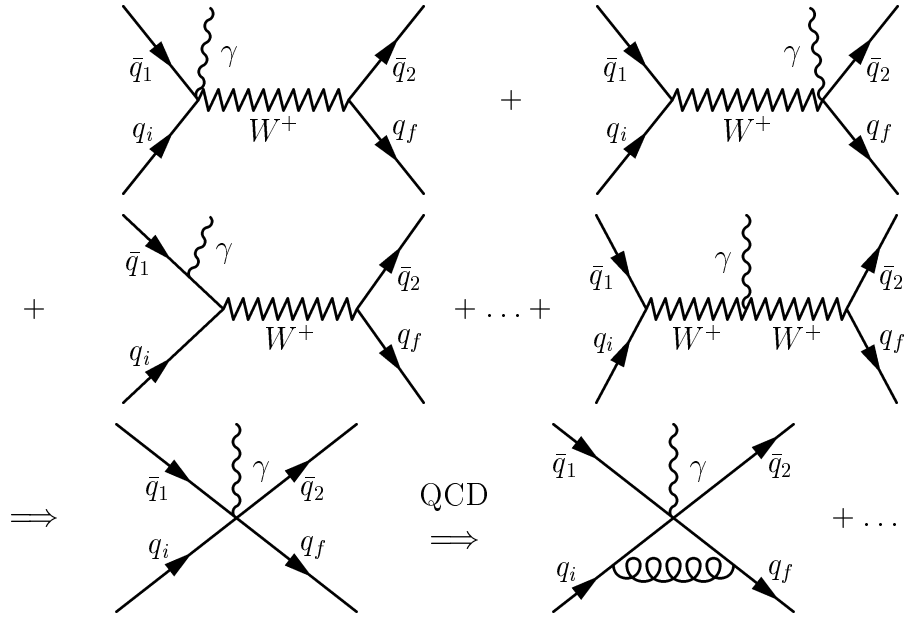}}
 \caption{Free quark Feynman diagrams representing the procedure of deduction of the point-like 
photon $\times$ current $\times$ current interaction Hamiltonian 
in the minimal NCSM. The diagram with double $W^+$ exchange is given for the sake of completeness,
since its contributions are suppressed by $G^2_F$ and consequently neglected. 
QCD corrections are indicated.
The different combinations of ${\bar q}_1$, ${\bar q}_2$, $q_i$ and $q_f$ produce different decay modes.
For example the combination $({\bar s}u) \to ({\bar d}u)\gamma$ 
represents the $K^+\to \pi^+ \gamma$ decay mode.}
 \label{fig9}
\end{figure}
The contributions to the $K\to \pi \gamma$ decay amplitude come from the Feynman diagrams given in Fig. \ref{fig9}.
The first two classes of diagrams in there, by
integrating out the heavy $W^+$-boson field,
effectively shrink into the fifth diagram, which represents in the momentum space, 
the effective, gauge-invariant, point-like, non-commutative 
photon $\times$ current $\times$ current interaction Hamiltonian \cite{dstw} 
in the static-quark approximation, responsible for
SM-forbidden $K(k)\to \pi(p)\gamma(q)$ decay:
\begin{eqnarray}
&&H_{NCSM}^{sqa}(Ajj^{\dagger})= i\;\frac{2{\sqrt 2}}{3}\;e\;G_F \;V^*_{us}V_{ud}^{} \;
\left(\epsilon_{\mu}(q)\;J^{\mu}(k,p)\right);
\label{4} \\
&&J^{\mu} (k,p)=q_{\nu}\left(\theta^{\mu\nu}j^{\alpha}
+\theta^{\nu\alpha}j^{\mu}+\theta^{\alpha\mu}j^{\nu}\right)j_{\alpha}^\dagger;\; q=k-p;
\nonumber \\
&&\left(\theta^{\mu\nu}j^{\alpha}+\theta^{\nu\alpha}j^{\mu}+\theta^{\alpha\mu}j^{\nu}\right)j_{\alpha}^\dagger
=\left({\bar\psi^s_L} \theta^{\mu\nu\alpha}\psi^u_L\right) \left( {\bar\psi^d_L} \gamma_{\alpha}{\psi^u_L} \right)^\dagger;
\nonumber \\
&& \theta^{\mu\nu\alpha}= 
\theta^{\mu\nu}\gamma^{\alpha}+\theta^{\nu\alpha}\gamma^{\mu}+\theta^{\alpha\mu}\gamma^{\nu},
\nonumber
\end{eqnarray}
where $\epsilon_{\mu}(q)$ is the photon polarization vector.
Note that in the calculation of the diagrams in Fig. \ref{fig9} we were using the valence quark approximation, 
i.e. the fact that quark--antiquark pairs in $K^+$ and $\pi^+$ mesons are collinear.

The flavour-changing parts of the charge current are defined as
\begin{equation}
j^{\mu}_L=\,\frac{1}{2}\,{\bar s}{\gamma^{\mu}}(1-\gamma_5)u ;
\:\:\:
j^{\mu\dagger}_L=\left(\,\frac{1}{2}\,{\bar d}{\gamma^{\mu}}(1-\gamma_5)u \right)^{\dagger}.
\label{6}
\end{equation}

Before proceeding to the next step of our calculations, we have to discuss the possible 
${\cal L}^{SM}$ + ${\cal L}^{NCSM}$ contributions that come
from the diagrams where the photon is attached to the quark and to the boson fields. 
Considering only vertices from SM and NCSM up to linear order in $\theta$, it is clear from diagrams in Fig. \ref{fig9}  
that we have to analyse altogether five diagrams. Vertices in diagrams are of the following type:
$j^{SM}_{nc}({\bar q}q\gamma)\,+\,j^{NCSM}_{nc}({\bar q}q\gamma)$, 
$j^{SM}_+({\bar q}qW)\,+\,j^{NCSM}_+({\bar q}qW)$,
and $j^{SM}_{nc}(WW\gamma)\,+\,j^{NCSM}_{nc}(WW\gamma)$.

First, the terms coming from the neutral currents (eq.(74) of Ref. \cite{cal}) 
are absent due to the static-quark approximation, i.e. diagrams where photons are attached to quark fields do
not contribute.
Second, isolating the $WWA$ terms from eq. (74) of Ref. \cite{cal}, we 
obtain a structure containing terms with power proportional to $eg^2$, the same as for the
pure SM diagram. However, integrating out heavy {\it W} fields \cite{dgh},
it is easy to see that diagrams contribute to the 
amplitude with power proportional to $eG^2_F$, and consequently we could safely neglect them. \\

The next important step is to introduce QCD effects, by considering gluon 
exchange contributions; see e.g. sixth diagram in Fig. \ref{fig9}. 
All the other contributions that originate from diagrams that contain vertices with more than two gauge
bosons (for example photon--photon--{\it W}) are of order $\theta^2$. We also note that a diagram with
a photon--gluon--gluon vertex does not exist in the minimal NCSM \cite{cal}.
Because of this QCD corrections to the NCSM photon$\times$current$\times$current Hamiltonian
are not affected by non-commutative
terms, i.e. they remain the same as in the case of the SM QCD enhanced effective weak Hamiltonian \cite{dgh}. 
This way, for the above current $\times$ current interactions, we have 
\begin{eqnarray}
\left(j^{\mu}_Lj^{\dagger}_{L\nu}\right)_{QCD}^{eff}
=\frac{1}{8}\left(c_-{\cal O}_- + c_+{\cal O}_+\right),
\end{eqnarray}
where the operators ${\cal O}_{\mp}$ are defined in the usual way \cite{dgh}:
\begin{eqnarray}
{\cal O}_{\mp} = {\bar s}^i{\gamma^{\mu}}(1-\gamma_5)u^i {\bar u}^j{\gamma_{\nu}}(1-\gamma_5) d^j
\mp {\bar s}^i{\gamma^{\mu}}(1-\gamma_5)d^i {\bar u}^j{\gamma_{\nu}}(1-\gamma_5) u^j,
\end{eqnarray}
with upper $i,j$ indices defining the colour quantum numbers. The one-loop corrections, i.e. the 
QCD enhancement (suppression) coefficients $c_-\,(c_+)$ at the renormalization scale $\mu \simeq 1$ GeV,
and $\Lambda_{QCD} \simeq 0.2$ GeV receive the following values $c_- \simeq 2.1$, $c_+ \simeq 0.4$.
Consequently, branching ratios receive an order of magnitude enhancement and/or suppression due to the QCD corrections.

Now we proceed with the calculation of the $K^+\rightarrow \pi^+ \gamma$ decay.
The hadronic matrix element $\langle \pi|jj^\dagger|K\rangle$ in the
vacuum saturation approximation has the following form:
\begin{eqnarray}
&& \langle \pi^+(p)|\left(j^{\mu}_Lj^{\dagger}_{L\nu}\right)^{eff}_{QCD}|K^+(k)\rangle
\nonumber\\
&&=\frac{1}{12}(c_-+2c_+)\langle \pi^+(p)|{\bar u}{\gamma^{\mu}}\gamma_5 d|0\rangle
\langle 0|{\bar s}{\gamma_{\nu}}\gamma_5u|K^+(k)\rangle
\nonumber \\
&&=\frac{1}{12}(c_-+2c_+)\left(-ip^{\mu}f_{\pi}\right)\left(ik_{\nu}f_K\right).
\label{10}
\end{eqnarray}

From the above expressions we found the amplitude for the $K^+\rightarrow\pi^+ \gamma$ decay (with $q=k-p$):
\begin{eqnarray}  
{\cal M}^{sqa}_{K\pi \gamma} &=& \frac{i}{3\sqrt 2}e\,G_F\;
V^*_{us}V^{}_{ud} f_{\pi}f_K \frac{1}{3}(c_-+2c_+)
\nonumber \\
&\times&
\epsilon_{\mu}(q)
\left[\widetilde{q}^{\mu}(pk)+p^{\mu}(q\theta k)-\widetilde{k}^{\mu}(pq)\right].
\label{11}
\end{eqnarray}

Taking the kaon at rest and performing the phase-space integrations, from 
the gauge-invariant amplitude ${\cal M}^{sqa}_{K\pi \gamma}$, 
\begin{eqnarray}
\sum_\mathrm{spins}\,|{\cal M}_{K\pi \gamma}|^2
&=&\frac{1}{18} e^2\,G^2_F|V^*_{us}V^{}_{ud}|^2 f_{\pi}^2f_K^2 \frac{1}{9}(c_-+2c_+)^2
\label{12}  \\
&\times&
\left[{\widetilde q}{\widetilde q}(pk)^2 -2{\widetilde q}{\widetilde k}(pk)(pq)+ 
{\widetilde k}{\widetilde k}(pq)^2  - (q\theta k)^2 p^2\right],
\nonumber 
\end{eqnarray}
we obtain the following expression for the branching ratio:
\begin{eqnarray}
&&BR^{sqa}(K^+\rightarrow\pi^+ \gamma)
=\tau_{K^+}\Gamma(K^+\rightarrow\pi^+ \gamma)
\label{15}\\
&&=\frac{{\tau_{K^+}}\alpha}{1728}G^2_F f^2_{\pi}f^2_K|V^*_{us}V^{}_{ud}|^2 \frac{1}{9}(c_-+2c_+)^2
\nonumber\\
&&\times\frac{m_K^5}{\Lambda^4_{NC}}\left(1-\frac{m^2_{\pi}}{m^2_K}\right)^3
\left[\left(1-\frac{m^2_{\pi}}{m^2_K}\right)^2 \sum_{i=1}^{3}(c^{0i})^2 +
\left(1+\frac{m^2_{\pi}}{m^2_K}\right)^2 \sum_{{i,j=1 \atop i<j}}^{3}(c^{ij})^2\right].
\nonumber
\end{eqnarray}

The QCD corrections turn out not to be of particular importance for our charged decay mode $K^+\to \pi^+\gamma$.
However, the neutral decay mode $K^0 \to \pi^0 \gamma$ is suppressed
by a factor of $(c_- -2c_+)^2/2(c_- +2c_+)^2$ relative to the charged one, owing to isospin and to the QCD corrections.\\

To maximize the branching ratio due to the effect of non-commutativity we assume that the square bracket in the above 
expression takes the value of 2. We are taking experimentay known quantities such as masses: $m_{\pi^+}$, $m_{K^+}$,
$m_{D^+}$ and $m_{B^+}$, mean lives: $\tau_{K^+}$, $\tau_{D^+}$ and $\tau_{B^+}$, CKM matrix elements:
$|V_{ud}|$, $|V_{us}|$ and $|V_{cd}|$, and pseudoscalar meson decay constants: $f_{\pi^+}$,
$f_{K^+}$ and  $f_{D_s^+}$ from the
Particle Data Group \cite{rpp}. We find the CKM matrix element $|V_{ub}|=0.0037$ in recently published BaBar results
\cite{babar}. Finally, we are using decay constants $f_{D^+}=215$ MeV and $f_{B^+}=186$ MeV from recent lattice calculations 
reported in Ref. \cite{smr}. The branching ratio for $K^+\to \pi^+\gamma$ as a function of the non-commutative scale 
$\Lambda_{NC}$ is:
\begin{eqnarray}
BR(K^+\rightarrow\pi^+ \gamma)\simeq 1.0\times10^{-5}\;\frac{(1\,\rm GeV^4)}{\Lambda^4_{NC}},
\end{eqnarray}
while the other interesting modes could easily be found from the following ratios:
\begin{eqnarray}
&&BR(K^+\to \pi^+\gamma):BR(D_s^+\to \pi^+\gamma):BR(D^+\to \pi^+\gamma):BR(B^+\to \pi^+\gamma)
\nonumber \\
&&\cong 1:2.40:0.20:0.01.
\end{eqnarray}
A very interesting mode is the $D^+_s \to \pi^+ \gamma$ decay, since it dominates
the other modes, 
because of the absence of the CKM suppression.
The branching ratios for $B^+\to (K^+,D^+) \gamma$ modes are very small.

For the non-commutativity scale of  $0.10,\;0.25,\;0.50,\;1.0$ TeV  we have found 
values of the branching ratio
$BR(K^+\rightarrow\pi^+ \gamma)\simeq 1.0\times10^{-13}$, $2.6\times10^{-15}$, 
$1.6\times10^{-16}$, and $1.0\times10^{-17}$, respectively.

All the above statements are of course true only in the static-quark approximation.\\

{\it Gauge invariance and the $K\to \pi \gamma$ decay in the SM}
\\
To show the correctness of our estimate of the $BR(K\to \pi\gamma)$ within the NCSM we will
next prove that the amplitude for $K\to \pi\gamma$ decay vanishes in the SM because of
the electromagnetic gauge condition.

There are two contributions to the decay amplitude $A(K\to \pi\gamma)|_{SM}$:\\

(P) the free quark amplitude arising from the 1-loop penguin diagrams 

\hspace{.5cm}
Fig. \ref{Zfig1}: $A^{peng.}(K\to \pi\gamma)|_{SM}$,\\

(T) the free quark amplitude coming out of tree diagrams 

\hspace{.5cm}
Fig. \ref{fig9}: $A^{tree}(K\to \pi\gamma)|_{SM}$, so that we have 
\begin{eqnarray}
A(K\to \pi\gamma)|_{SM}=A^{peng.}(K\to \pi\gamma)|_{SM}+A^{tree}(K\to \pi\gamma)|_{SM}
\end{eqnarray}

The proof that $A(K\to \pi\gamma)|_{SM}=0$ proceeds in the following steps.\\
(1) We write the SM penguin contributions to the free quark amplitude.\\
(2) The five free quark diagrams with a photon coming out of quark legs and the photon out of the W propagator, 
from Fig. \ref{fig9}, contribute to the SM tree amplitude. We estimate those diagrams 
in the 't Hooft--Feynman gauge using the standard argument for the Feynman propagator,
\begin{eqnarray}
 \int d^4x \Delta_F^{\mu\nu}(x,m_W)=
 \int d^4x\,d^4k\,\frac{1}{(2\pi)^4}\frac{-g^{\mu\nu}+\frac{k^{\mu}k^{\nu}}{m^2_W}}{k^2-m^2_W} e^{-ikx}=\frac{g^{\mu\nu}}{m^2_W}.
\end{eqnarray}
(3) Next we hadronize the SM free quark amplitudes by sandwiching the interaction (four-quark) operator, 
between the time-independent state-vectors $\langle \pi^+|$ and $|K^+\rangle$.
This corresponds to the well known Heisenberg picture \cite{wein}.\\
(4) We apply Lorentz decomposition of the relevant hadronic matrix element in the penguin amplitude 
and use the vacuum saturation approximation and PCAC in the tree amplitude evaluations.\\
(5) We assume that the meson is described within the valence quark approximation and that 
quark and antiquark are collinear, each carrying a half of the meson momenta.\\

(P) From Fig. \ref{Zfig1}, i.e. from the first equation in section 1.2 for a real photon we have 
\begin{eqnarray}
&&A^{peng.}(K\to \pi\gamma)|_{SM}
\\
&&=iG_2\;\langle \pi^+(p)|\left( m_d{\bar d}\sigma_{\mu\nu}q^{\nu}s_L + 
m_s{\bar d}\sigma_{\mu\nu}q^{\nu}s_R\right)|K^+(k)\rangle \;\epsilon^{\mu}(q).
\nonumber
\end{eqnarray}
Next we use the Lorentz decomposition of the $\sigma_{\mu\nu}$ operator matrix element and find
\begin{eqnarray}
\langle \pi^+(p)|{\bar d}\sigma_{\mu\nu}q^{\nu}s|K^+(k)\rangle \;\epsilon^{\mu}(q)
&=&\left(k_{\mu}p_{\nu}-k_{\nu}p_{\mu}\right)\;q^{\nu}\;\epsilon^{\mu}(q) \;f(q^2)
\\
&= &(kq)\;(q_{\mu}\epsilon^{\mu}(q))\;f(q^2)=0,
\nonumber
\end{eqnarray}
which means that $A^{peng.}(K\to \pi\gamma)|_{SM}=0$.\\

(T) We start to calculate the diagram 
with a photon coming out of the W propagator, Fig. \ref{fig9}. After a trivial integrations 
over the delta functions, the amplitude reads
\begin{eqnarray}
&&A^{tree}_W|_{SM}= ie\frac{g^2}{4}V^*_{us}V^{}_{ud}
{\bar\psi}_{u_f}(p_{u_f})\gamma_{\alpha}(1-\gamma_5)\psi_{\bar d}(p_{\bar d})
\frac{g^{\alpha\tau}}{m^2_W}
\nonumber\\
&&\times \left[(q-p)_{\nu}g_{\mu\tau}+(p-k)_{\mu}g_{\tau\nu}+(k-q)_{\tau}g_{\mu\nu}\right]
\frac{g^{\nu\beta}}{m^2_W}
\nonumber\\
&&\times {\bar\psi}_{\bar s}(p_{\bar s})\gamma_{\beta}(1-\gamma_5)\psi_{u_i}(p_{u_i})\;\epsilon^{\mu}(q).
\end{eqnarray}
Momentum conservation: $k=p_{\bar s}+p_{u_i}$, $p=p_{\bar d}+p_{u_f}$, $k=p+q$,
and the assumptions (3)--(5) gives: 
\begin{eqnarray}
\langle A^{tree}_W\rangle_{SM}=ie{\sqrt2}G_F\,f_{\pi}f_K\,V^*_{us}V^{}_{ud}
\frac{kq}{m^2_W}\;\left(q_{\mu}\epsilon^{\mu}(q)\right)=0.
\end{eqnarray}

Next, we estimate the free-quark amplitude from the diagram where the photon is 
coming out of the antiquark $\bar s$ leg, Fig. \ref{fig9}. After a trivial integration we found
\begin{eqnarray}
A^{tree}_{\bar s}|_{SM}&=& i\frac{e}{3}\frac{g^2}{4} V^*_{us}V^{}_{ud}
{\bar\psi}_{u_f}(p_{u_f})\gamma_{\nu}(1-\gamma_5)\psi_{\bar d}(p_{\bar d})
\frac{g^{\nu\tau}}{m^2_W}
\nonumber\\
&\times& {\bar\psi}_{\bar s}(p_{\bar s})\gamma_{\tau}
\frac{({\not \!p_{\bar s}}-{\not \!q})+m_s}{(p_{\bar s}-q)^2-m^2_s}
\gamma_{\mu}(1-\gamma_5)\psi_{u_i}(p_{u_i})\;\epsilon^{\mu}(q).
\end{eqnarray}
Using the assumption (3)--(5), from the above denominators we obtain a factor $1/(kq)m^2_W$.
Using Dirac algebra identities to reduce 
$\gamma_{\nu}({\not \!p_{\bar s}}-{\not \!q})\gamma_{\mu}$ term,
and assumptions (4), with the help of definition $kq=pq=(k^2-p^2)/2$ 
${\bar\psi}_L^{\bar s}\gamma_{\nu}\gamma_{\mu}\psi_L^{u_i} =0$,
we obtain the following amplitude 

\begin{eqnarray}
\langle A^{tree}_{\bar s}\rangle_{SM}=i\frac{-1}{3}e{\sqrt2}G_F\,f_{\pi}f_K\,V^*_{us}V^{}_{ud}\,
\frac{kp}{kq}(k_{\mu}\epsilon^{\mu}(q)).
\end{eqnarray}
The amplitude coming from the second initial leg is:
\begin{eqnarray}
\langle A^{tree}_{u_i}\rangle_{SM}=2\langle A^{tree}_{\bar s}\rangle_{SM}.
\end{eqnarray}
The amplitudes from the outgoing quark--antiquark pair are
\begin{eqnarray}
\langle A^{tree}_{u_f}\rangle_{SM}&=&2\langle A^{tree}_{\bar d}\rangle_{SM}
\\
&=&i\frac{2}{3}e{\sqrt2}G_F\,f_{\pi}f_K\,V^*_{us}V^{}_{ud}\,\frac{kp}{pq}(p_{\mu}\epsilon^{\mu}(q)).
\nonumber
\end{eqnarray}
Summing up the above four contributions, we have
\begin{eqnarray}
\left(\langle A_{u_i}\rangle +\langle A_{\bar s}\rangle +\langle A_{u_f}\rangle +\langle A_{\bar d}\rangle\right)^{tree}_{SM}
=-ie{\sqrt2}G_Ff_{\pi}f_KV^*_{us}V^{}_{ud}
\frac{kp}{kq}\,(q_{\mu}\epsilon^{\mu}(q))=0,
\nonumber
\end{eqnarray}
which finally gives
\begin{eqnarray}
A^{tree}(K\to \pi\gamma)|_{SM}=\left(\langle A_W\rangle +\langle A_{u_i}\rangle +
\langle A_{\bar s}\rangle +\langle A_{u_f}\rangle +\langle A_{\bar d}\rangle\right)^{tree}_{SM}=0.
\end{eqnarray}
By this, we prove our statement that the amplitude for $K\to \pi\gamma$ decay vanishes in the SM, 
because of the electromagnetic gauge condition, i.e. 
\begin{eqnarray}
A(K\to \pi\gamma)|_{SM}=A^{peng.}(K\to \pi\gamma)|_{SM}+A^{tree}(K\to \pi\gamma)|_{SM}=0.
\end{eqnarray}

\subsection{\it Non-commutative Abelian gauge theories}

In the last part of these lectures we discuss a possible mechanism
for additional energy loss in stars induced by space-time 
non-commutativity. The mechanism is based on neutrino--antineutrino
coupling to photons, which arises quite naturally in non-commutative
Abelian gauge theory \cite{rstw}.

We are interested in an effective model of 
particle physics involving neutrinos and photons
on non-commutative space-time. More specifically we need to
describe the scattering of particles 
that enter from an asymptotically commutative
region into a non-commutative interaction region.
We shall focus on a model that satisfies the following requirements:
\begin{enumerate}
\item[(i)] Non-commutative effects are described perturbatively. The action is
written in terms of assymptotic commutative fields. 
\item[(ii)] The action is gauge-invariant under $U(1)$-gauge transformations.
\item[(iii)] It is possible to extend the model to  a non-commutative electroweak
model based on the gauge group $U(1) \times SU(2)$.
\end{enumerate}
As we have already argued in these lectures the action of such an effective model
differs from  the commutative theory essentially by the presence of star products
and Seiberg--Witten maps. The Seiberg--Witten maps are necessary to express
the non-commutative fields $\hat \psi$, $\hat A_\mu$ that appear in the action
and that transform under non-commutative gauge transformations, in terms
of their asymptotic commutative counterparts $\psi$ and $A_\mu$.
The coupling of matter fields to Abelian gauge bosons is a non-commutative
analogue of the usual minimal coupling scheme. Neutrinos do not carry a $U(1)$
(electromagnetic) charge and hence do not directly couple to Abelian
gauge bosons (photons) in a commutative setting. In the presence of space-time
non-commutativity, it is, however, possible to couple neutral particles to 
gauge bosons via a star commutator. The relevant covariant derivative is
\begin{equation}
\hat D_\mu \hat \psi = \partial_\mu \hat \psi - i e \hat A_\mu \star \psi
+ i e \hat\psi \star \hat A_\mu \; ,
\end{equation}
with a coupling constant $e$. Here one may think of the non-commutative
neutrino field $\hat \psi$ as having left charge $+e$, right charge $-e$
and total charge zero. From the perspective of non-Abelian gauge theory,
one could also say that the neutrino field is charged in a non-commutative
analogue of the adjoint representation. Physically such a coupling of 
neutral particles to gauge bosons is possible because the non-commutative
background is described by an antisymmetric tensor $\theta^{\mu\nu}$
that plays the role of an external field in the theory. The photons
do not directly couple to the ``bare'' commutative neutrino fields,
but rather modify the non-commutative background.
The neutrinos propagate in that background.\\

The action for a neutral fermion that couples to an 
Abelian gauge boson in a non-commutative background is \cite{rstw}:
\begin{equation}
S = \int d^4 x \left(\,\bar{\widehat \psi} \star i\gamma^\mu\widehat D_\mu \widehat\psi
-m \bar{\widehat \psi} \star \widehat\psi\right).
\label{1}
\end{equation}
Here $\widehat \psi = \psi +e \theta^{\nu\rho} A_\rho \pp\nu \psi + \mathcal{O}(\theta^2)$ and 
$\widehat A_\mu = A_\mu + \theta^{\rho\nu}A_{\nu}
\left[\partial_{\rho}A_{\mu}-\frac{1}{2}\partial_{\mu}A_{\rho}\right]+ \mathcal{O}(\theta^2)$ 
is the Abelian NC gauge potential expanded by the SW map. 

To first order in $\theta$, the action reads
\begin{eqnarray}
&&S = \int d^4 x \, \left\{ \bar \psi 
\left[i\gamma^\mu \pp\mu  - m\left(1+e\theta^{\mu\nu}A_{\mu\nu}\right)\right]\psi
\right. \label{2}\\
&&\left. 
+ ie \theta^{\mu\nu} \left[(\pp\mu \bar \psi) A_\nu \gamma^\rho (\pp\rho \psi)
-  (\pp\rho\bar \psi) A_\nu \gamma^\rho  (\pp\mu  \psi)
+  \bar \psi (\pp\mu A_\rho) \gamma^\rho  (\pp\nu \psi) \right]\right\}.
\nonumber
\end{eqnarray}
Integrating by parts, this can also be written in a manifestly gauge-invariant way as 
\begin{eqnarray}
S = \int d^4 x  {\bar{\psi}} \left[i\gamma^{\mu}\pp\mu 
- m\left(1+e\theta^{\mu\nu}A_{\mu\nu}\right)
- ieA_{\mu\nu}\left(\frac{1}{2}\theta^{\mu\nu}{\gamma^{\rho}\pp\rho}+ 
\theta^{\nu\rho} \gamma^{\mu}\pp\rho\right)\right]\psi.
\nonumber
\end{eqnarray}
The above action represents the tree-level point-like interaction of the photon and neutrinos.
We could also call it ``the background field anomalous-contact'' interaction.

\subsubsection{\it The plasmon decay to neutrino--antineutrino pairs}

To obtain the ``transverse plasmon'' decays in the stars on the scale of non-commutativity, we start
with the action determining the $\gamma\nu\bar\nu$ interaction. In a stellar
plasma, the dispersion relation of photons is identical with that of 
a massive particle \cite{Jancovici}--\cite{Salati}
\begin{equation}
q^2 \equiv {\rm E}_\gamma^2-{\bf q}_\gamma^2=\omega_{\rm pl}^2
\end{equation}
with $\omega_{\rm pl}$ being the plasma frequency. \\

From Eq. (\ref{2}) we extract, for the left--right massive neutrinos, the following Feynman rule for  
the ${\gamma}(q)\to {\nu}(k'){\bar {\nu}}(k)$ vertex in momentum space:
\begin{equation}
{\Gamma}^{\mu}_{{\rm L} \choose {\rm R}}({\nu}{\bar {\nu}}{\gamma})
=ie\frac{1}{2}(1 \mp \gamma_5)
\left[(q\theta k)\gamma^{\mu}+({\not \!k}-2m_{\nu}){\widetilde q}^{\mu}-{\not \!q}{\widetilde k}^{\mu}\right].
\label{5}
\end{equation}
In the case of massless neutrinos the Feynman rule reads:
\begin{equation}
{\Gamma}^{\mu}_{{\rm L} \choose {\rm R}}({\nu}{\bar {\nu}}{\gamma})
=ie\frac{1}{2}(1 \mp \gamma_5){\theta}^{\mu\nu\tau}k_{\nu}q_{\tau},\;\;
\label{4}
{\theta}^{\mu\nu\tau}={\theta}^{\mu\nu}\gamma^{\tau}+{\theta}^{\nu\tau}\gamma^{\mu}+{\theta}^{\tau\mu}\gamma^{\nu}.
\nonumber
\end{equation}
Here $q_{\mu}\Gamma^{\mu}=0$ explicitly shows the electromagnetic gauge invariance of the above vertices.\\

From the gauge-invariant amplitude ${\cal M}_{\gamma {\nu} {\bar{\nu}}}$ in momentum space 
for plasmon (off-shell photon) decay to the
left and/or right massive neutrinos in the NCQED, we find: 
\begin{eqnarray}
\sum_{\rm pol.} |{\cal M}_{\gamma {\nu} {\bar{\nu}}}|^2 =
 4e^2\left[\left(q^2 -2m_{\nu}^2\right)\left(\frac{5}{2}m_{\nu}^2{\widetilde q}^2-(q\theta k)^2\right) 
 +m_{\nu}^2 q^2 ({\widetilde k}^2-{\widetilde k}{\widetilde q})\right].
\end{eqnarray}
In the rest frame of plasmon-medium we have
\begin{eqnarray}
{\widetilde q}^2={\rm E}_{\gamma}^2 \sum_{i=1}^3 (\theta^{0i})^2
=\frac{{\rm E}_{\gamma}^2}{\Lambda^4_{\rm NC}}\sum_{i=1}^3 (c^{0i})^2 
\equiv \frac{{\rm E}_{\gamma}^2}{\Lambda^4_{\rm NC}}{\vec E}^2_{\theta}\equiv q \theta^2 q,
\end{eqnarray}
from where we then find \cite{rstw}:
\begin{eqnarray}
&&\Gamma(\gamma_{\rm pl} \rightarrow {\bar{\nu}}_{{\rm L}\choose {\rm R}}\nu_{{\rm L}\choose {\rm R}})
= \frac{\alpha}{48}\frac{\omega^6_{\rm pl}}{{\rm E}_{\gamma}\Lambda^4_{\rm NC}}\sqrt{1-4\frac{m_{\nu}^2}{\omega_{\rm pl}^2}}\\
&&\times \left[\left(1+20\frac{m^2_{\nu}}{\omega^2_{\rm pl}}-48\frac{m^4_{\nu}}{\omega^4_{\rm pl}}\right)\sum_{i=1}^{3}(c^{0i})^2 + 
2\frac{m^2_{\nu}}{\omega^2_{\rm pl}}\left(1-4\frac{m^2_{\nu}}{\omega^2_{\rm pl}}\right)\sum_{{i,j=1 \atop i<j}}^{3}(c^{ij})^2\right].
\nonumber
\end{eqnarray}
In the all above calculations we have used the notation:
\begin{eqnarray}
&&{\widetilde q}^2=|{\theta^{\mu\nu}}q_{\nu}|^2
=({\theta^{\mu\nu}}q_{\nu})({\theta_{\mu\rho}}q^{\rho})^{\dagger}
=-({\theta^{\mu\nu}}q_{\nu})(\theta_{\mu\rho}q^{\rho}),
\\
&&{\theta}^2= {\theta }^{\mu\nu}{\theta }_{\nu\mu}=({\theta}^2)^{\mu}_{\mu}
=\frac{2}{\Lambda^4_{\rm NC}}\left(\sum_{i=1}^{3}(c^{0i})^2 - \sum_{{i,j=1 \atop i<j}}^{3}(c^{ij})^2\right)
\equiv \frac{2}{\Lambda^4_{\rm NC}}\left({\vec E}^2_{\theta}-{\vec B}^2_{\theta}\right).
\nonumber
\end{eqnarray}
In the above expression we
parametrize the $c_{0i}$'s by introducing the angles characterizing the background ${\theta}^{\mu\nu}$ 
field of the theory:
\begin{eqnarray}
c_{01}=\cos\xi,\;\;c_{02}=\sin\xi \;\cos\zeta,\;\;c_{03}=\sin\xi \;\sin\zeta,
\end{eqnarray}
where $\xi$ is the angle between the ${\vec E}_{\theta}$ field and the direction of the incident beam,
i.e. the photon axes. The angle $\zeta$ defines the origin of the $\phi$ axis.  
The $c_{0i}$'s are not independent; in pulling out the overall scale $\Lambda_{\rm NC}$ we can always
impose the constraint $\sum_{i=1}^3 (c^{0i})^2=1$. Here we consider three physical cases:
$\xi=0,\;\pi/4,\;\pi/2$, which for $\zeta = \pi/2$ satisfy the imposed constraint.
 This parametrization provides a good physical interpretation of the NC effects.\\

In the rest frame of the medium, the decay rate of a ``transverse plasmon'',
of energy ${\rm E}_\gamma$ and for the left--left and/or right--right massless neutrinos, is given by
\begin{equation}
\Gamma_{\rm NC}
(\gamma_{\rm pl}\to \nu_{{\rm L}\choose {\rm R}}\bar\nu_{{\rm L}\choose{\rm R}})
=\frac{\alpha}{48}\,\frac{1}{\Lambda_{\rm NC}^4}\,\frac{\omega_{\rm pl}^6}{{\rm E}_\gamma}.
\end{equation}

The Standard Model (SM) photon--neutrino interaction at tree
level does not exist. However, the effective photon--neutrino--neutrino vertex 
$\Gamma^{\mu}_{\rm eff}(\gamma\nu\bar\nu)$ is
generated through 1-loop diagrams, which are very well known in heavy-quark physics as ``penguin''
diagrams. Such effective interactions \cite{dsm,as} give non-zero charge radius, as well as the
contribution to the ``transverse plasmon'' decay rate. For details see Ref. \cite{as}.  
Finally, note that the dipole moment operator
$\sim em_{\nu}G_{\rm F}{\bar\psi}_{\nu}\sigma_{\mu\nu}\psi_{\nu}A^{\mu\nu}$, 
also generated by the ``neutrino-penguin'' diagram, 
gives negligible contributions because of the smallness of the neutrino mass, 
i.e. $m_{\nu} < 1$ eV \cite{nobel}.
The corresponding SM result is \cite{as}
\begin{equation}
\Gamma_{\rm SM}\left(\gamma_{\rm pl}\to {\nu_{\rm L}}{\bar\nu}_{\rm L}\right)
=\frac{{\rm c}_{\rm v}^2 G_{\rm F}^2}{48\pi^2\alpha }\;\frac{\omega_{\rm pl}^6}{{\rm E}_\gamma}.
\end{equation}
For $\nu_e$ we have ${\rm c}_{\rm v}=\frac{1}{2}+2\sin^2\Theta_{\rm W}$
while for $\nu_\mu$ and $\nu_\tau$ we have
${\rm c}_{\rm v}=-\frac{1}{2}+2\sin^2\Theta_{\rm W}$.  Comparing the
decay rates into all three left-handed neutrino families we thus need
to include a factor of~3 for the NC result, while 
${\rm c}_{\rm v}^2 \cong 0.8$ for the SM result \cite{rpp}. Therefore, the ratio of the rates is
\begin{equation}
\Re\equiv\frac{\sum_{\rm flavours}
\Gamma_{\rm NC}(\gamma_{\rm pl}\to {\nu_L}{\bar\nu}_{\rm L} + {\nu_{\rm R}}
{\bar\nu}_{\rm R})}
{\sum_{\rm flavours}
\Gamma_{\rm SM}(\gamma_{\rm pl}\to {\nu_{\rm L}}{\bar\nu}_{\rm L})}
=\frac{6\pi^2\alpha^2}{{\rm c}_{\rm v}^2G_{\rm F}^2\Lambda_{\rm NC}^4}.
\end{equation}
A standard argument involving globular cluster stars tells us that any
new energy-loss mechanism must not exceed the standard neutrino losses
by much, see section 3.1 in Ref. \cite{raffelt}.
Put another way, we should approximately
require $\Re<1$, translating into
\begin{equation}
\Lambda_{\rm NC}>\left(\frac{6\pi^2\alpha^2}{{\rm c}_{\rm v}^2 G_{\rm F}^2}\right)^{1/4}
\approx 81~{\rm GeV}\,.
\end{equation}
In the case of the absence of the sterile neutrinos ($\nu_{\rm R}$) in globular cluster stars
the scale of non-commutativity is approximately $\Lambda_{\rm NC}> 68~{\rm GeV}$.

\subsection{\it Discussion and conclusions on forbidden decays}

At the beginning of our discussion and conclusions, a very important comment is in order.

Extreme care has to be taken when one tries to compute matrix elements in NCGFT. In our model, the 
\emph{in} and \emph{out} states can be taken to be ordinary \emph{commutative} particles.
Quantization is straightforward to the order in $\theta$ that we have considered;
the Feynman rules can be obtained either through the Hamiltonian formulation or directly from the
Lagrangian; a rather convenient property of the action, relevant to computations, is
its symmetry under ordinary gauge transformations, in addition to non-commutative ones.\\

We propose decay modes that are strictly SM-forbidden, namely
$Z \rightarrow \gamma\gamma$, $K\to \pi\gamma$, ..., as a possible signature of non-commutativity. 
An experimental discovery of $Z \rightarrow \gamma\gamma$, $K\to \pi\gamma$, ...,
decays would certainly indicate a violation of the accepted SM and the definite appearance
of new physics. To determine whether such SM breaking is ultimatlly coming from
space-time non-commutativity or from some other source would require a tremendous amount
of additional theoretical and experimental work, and is beyond the scope of the present work.\\

The structure of our main results for the gauge sector, (\ref{eqn0}) to (\ref{eqn3}), remains the same
for  $SU(5)$ and $SU(3)_C \times SU(3)_L \times SU(3)_R$ GUTs that
embed the NCSM that is based on the SW map \cite{ASCH,Desh}; only the coupling
constants change. In the particular case of $SO(10)$ GUTs
there is no triple gauge boson coupling \cite{ASCH}.
This is due to the same Lorentz structure of 
the gauge boson couplings  ${Z\gamma\gamma}$ and ${Zgg}$ in our NCSM and 
in the above GUTs, understood underlying theories for the NCSM.
In the GUT framework, the triple-gauge couplings could be uniquely fixed.
However, the GUT couplings have to be evolved down
to the TeV scale. This requires additional theoretical work, 
and it is a subject for another study.

Note finally that the inclusion of other triple-gauge boson interactions in $2\to 2$ experiments
sufficiently reduce available parameter space of our model. This way it is possible to fix all the 
coupling constants from the NC gauge sector.

To get some idea of the values, let us choose the central value of the $Z\gamma\gamma$
coupling constants ${|{\rm K}_{Z\gamma \gamma}|\simeq 0.1}$ and assume that maximal non-commutativity
occurs at the scale of $\sim$1 TeV. The resulting branching ratio for our $Z\to \gamma \gamma$ decay would 
then be ${\cal O}(10^{-8})$, which is a reasonable order of magnitude. \\

The dynamics of the SM forbidden flavour changing weak decays is described in the framework of 
the so-called minimal NCSM developed by the Wess group \cite{cal}.
The branching ratios are roughly estimated within the static-quark approximation. 
Despite the simplifications gained by the static-quark approximation, we did obtain reasonable 
results, i.e. expected rates. Namely, in the static-quark approximation many terms did not contribute at all.
An improved estimate, by inclusion of all those terms, would certainly increase our branching ratios. 
We do expect increasing to more than one order of magnitude, which would than place the
$BR(K^+\rightarrow\pi^+ \gamma)$ closeer to today's experimentally accessible range \cite{litt,litte}.

The same increase should also take place for the $B \to K \gamma$ modes via 1-loop
non-commutative FCNC, i.e. via non-commutative penguin diagrams \cite{iltan}. 
Namely we know that penguin diagrams, in the case of B-meson decays, have a number of advantages over the tree 
diagrams. Also the whole B sector has advantages over the kaon sector:\\

(a) rate is proportional to $m_B^5$ which cancels small mean life 
$\tau_B$ and small CKM matrix elements relative to kaons, i.e. 
\begin{equation}
\frac{({\tau_B\;m_B^5|V^{*}_{ts}V^{}_{tb}|^2})_{\rm peng.}}{({\tau_K\;m_K^5 |V^{*}_{us}V^{}_{ud}|^2})_{\rm tree}} \simeq 1;
\end{equation}

(b) penguins do not suffer from relatively small CKM matrix elements;\\

\begin{figure}
 \resizebox{0.9\textwidth}{!}{%
  \includegraphics{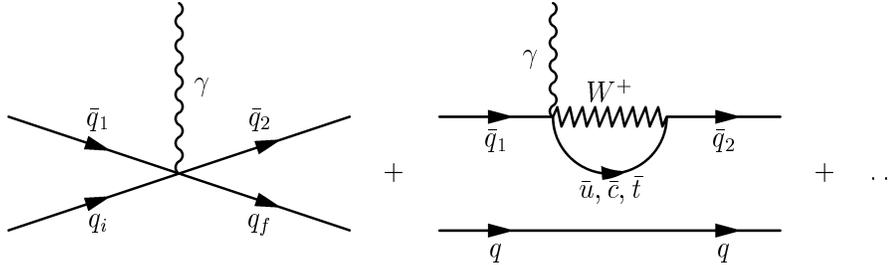}}
 \caption{The NCSM linear $\theta$-dependent contributions to typical flavour-changing diagrams. 
 The first one arises from
 the point-like charged current NCSM interactions, see for instance Fig. \ref{fig9},
 while the second represents the flavour-changing NCSM neutral-current, 1-loop transitions, 
 i.e. the typical non-commutative penguin diagram. 
 The ${\bar q}_1$, ${\bar q}_2$, $ q_i$ and $q_f$ are the same as in Fig. 9.}
 \label{fig10}
\end{figure}

(c) in the non-commutative penguin diagrams from the charm and top loops, Fig. \ref{fig10}, 
one might expect large QCD effect, i.e. the logarithmic type, $\sim[\alpha_s{\rm ln}({m_t^2}/{m_c^2})]$,
of the rate enhancement;\\

(d) note, however that the calculation of the non-commutative penguin diagrams would be highly complicated,
and would require a number of additional studies, to deal in particular with UV and/or IR divergences.
There already is  a lot in the literature  concerning the problem of (non-)renormalizability of
the non-commutative gauge field theories \cite{UV/IR}.\\

From the advantages described in (a) to (d), we conclude that some particular decay modes within 
the kaon and/or B meson sectors would receive 
the contributions from non-commutative tree and from non-commutative penguin diagrams of comparable size.
This is very important for the experimentalists, since it shows implicitly that some decay modes could be
relatively large, that means closer than we expect to the experimentaly accessible range.\\

The limit on the scale of non-commutativity from the energy loss in stars
depends on the requirement $\Re<1$ and from that point of view,
the constraint $\Lambda_{\rm NC}> 80~{\rm GeV}$,
obtained from the energy loss in the globular stellar clusters,
represents the lower bound on the scale of non-commutative gauge field theories.\\

Concerning the forbidden decays, the experimental situation can be summarized as follows:\\

(1) The joint effort of the DELPHI, ALEPH, OPAL and L3 Collaborations \cite{LEPEWWG} give us a hope that in 
not to much time all collected data from the LEP experiments will be counted and analysed, producing 
tighter bounds on triple-gauge boson couplings.
Finally, note that the best testing ground for studies of anomalous triple-gauge boson couplings, 
before the start of the linear $e^+e^-$ collider there will be the LHC. See for instance Ref. \cite{muller}.\\

(2) The authors of Brookhaven Experiment E787 recently published a new upper limit on the branching ratio
$BR(K^+\rightarrow\pi^+ \gamma)<3.6 \times 10^{-7}$ \cite{litt}. The E787 has been upgraded to a 
more sensitive experiment, E949, curently under way at the AGS. In this experiment it would be possible to 
the push sensitivity to $K^+\rightarrow\pi^+ \gamma$ by a quite large 
factor if there were sufficient motivation to do so \cite{litte}. 
We hope that the results of this research will convince the E949 Collaboration to go for it. \\

(3) In the future machines the productions of $10^{12}$, $10^{13}$, and $10^{14}$,
${\bar B}B$, ${\bar D}D$, and ${\bar K}K$ pairs  is expected , respectively.\\

(4) The sensitivity to the NC parameter $\theta^{\mu\nu}$ could be in the range of the next
generation of linear colliders, with a c.m.e. around a few TeV. \\

(5) We hope that, in the near future,
more sophisticated methods to observe, and more accurate techniques 
to measure the energy loss in the
stellar clusters will produce more restricting limits to the requirement $\Re<1$, 
something like $\Re<1/10$, and consequently a firmer 
bound on the scale of non-commutativity $\Lambda_{\rm NC}$.\\ 

In conclusion, both the hadron and the gauge sector of the NCSM as well as 
the NCQED are excellent places to discover
space-time non-commutativity experimentally.
We believe that the importance of a possible
discovery of non-commutativity of space-time at very short distances would convince
particle and astroparticle physics experimentalists to look for SM-forbidden decays in those sectors.
\\
\vspace{.2cm}
\\
I would like to thank for helpful discussions to L. Alvarez-Gaume, A. Armoni, N.G. Deshpande, G. Duplan\v ci\' c, R. Fleischer,
T. Hurth, Th. M\" uller, M. Prasza{\l}owicz, G. Raffelt, V. Ruhlmann-Kleider, P. Schupp and J. Wess.\\
This work was supported by the Ministry of Science and Technology of Croatia under Contract No. 0098002.


\begin{thebibliography}{99}
\bibitem{dstw}  The complet calculations will soon be published.
\end{thebibliography}
\end{document}